\documentclass{article}
\usepackage{iclr2027_conference,times}

\usepackage{algorithm}
\usepackage{algpseudocode}
\usepackage{float}
\usepackage[utf8]{inputenc}
\usepackage[T1]{fontenc}
\definecolor{darkblue}{rgb}{0.0, 0.0, 0.55}
\usepackage[colorlinks=true,
            linkcolor=darkblue,
            urlcolor=darkblue,
            citecolor=darkblue,
            anchorcolor=darkblue]{hyperref}
\usepackage{url}
\usepackage{xurl}
\newcommand{\doi}[1]{\url{https://doi.org/#1}}
\usepackage{booktabs}
\usepackage{amsfonts}
\usepackage{nicefrac}
\usepackage{microtype}
\usepackage{xcolor}
\usepackage{amsmath}
\usepackage{graphicx}
\usepackage{longtable}
\usepackage[normalem]{ulem}
\usepackage{pifont}
\usepackage{colortbl}

\newcommand\msmall[1]{\mbox{\scriptsize\ensuremath{#1}}}

\title{GEODE: Symmetry-Preserving Cartesian Diffusion for Crystal Generation}

\author{Yuchen Lou\thanks{Co-corresponding authors.}\phantom{\textsuperscript{\normalfont$*$}} \& Alex M. Ganose\footnotemark[1] \\
Department of Chemistry, Molecular Sciences Research Hub \\
Imperial College London, White City Campus \\
Wood Lane, London W12 0BZ, UK \\
\texttt{\{yuchen.lou19,a.ganose\}@imperial.ac.uk}
}

\iclrfinalcopy 
\begin{document}
\maketitle
\lhead{Preprint}

\begin{abstract}
Most known inorganic crystals exhibit symmetric atomic arrangements, yet generative models often fail to reproduce them.
Explicitly enforcing these symmetries has so far yielded fewer stable and novel structures than unconstrained generation.
We introduce Generative Equivariant Orbit Diffusion Engine (GEODE), to our knowledge the first model to combine coordinate and lattice diffusion in Cartesian space.
GEODE first samples symmetry templates, then jointly generates the lattice, atomic coordinates and atom types while preserving the specified symmetry with a novel Wyckoff-constrained loss.
Cartesian diffusion gives coordinate noise a consistent physical scale that we empirically demonstrate improves performance.
Unconditional generation achieves a metastable, unique and novel (mSUN) rate of 11.9\%, compared with 7.7\% for the next best symmetry-aware model.
We also introduce sampling time template filtering, which increases mSUN by $\sim$6\% without retraining, making GEODE competitive with leading symmetry-agnostic models.
Template selection also enables joint symmetry and property guidance, which we demonstrate through classifier-free guidance of permittivity.

\end{abstract}

\section{Introduction}
\label{sec:intro}

The established approach to computational materials discovery is high-throughput screening, where a pool of candidates is filtered against property criteria to identify promising materials \citep{jain2013materials, curtarolo2013high}.
However, screening can only identify materials included in existing databases \citep{park2024generative}.
Generative models learn to sample new structures directly instead of filtering existing ones.
A crystal is described by its lattice, atomic coordinates and atom types, all of which must be generated together.
As with molecules, the physical system is invariant under $E(3)$ transformations, but crystals also require the model to account for periodicity of lattice \citep{hahn1983international}.

A crystal's space group is the set of symmetry operations, such as translations and rotations, that leave it unchanged.
Most experimentally known crystals have non-trivial space group symmetry, with atoms on special Wyckoff positions constrained by these operations \citep{levy2025symmcd, chang2025sgequidiff}.
Existing generative models learn from databases which contain a broad range of space groups \citep{jain2013materials,schmidt2024alexandria} but disproportionately generate structures in the trivial space group $P1$ \citep{levy2025symmcd}.
MatterGen \citep{zeni2025mattergen} addresses this bias using classifier-free guidance \citep{ho2022cfg} on space-group labels to guide towards a target symmetry.
However, it only recovers the target space group 10--20\% of the time \citep{levy2025symmcd, chang2025sgequidiff}.
We argue that space group symmetry should be built into the model architecture rather than supplied as a conditioning label.
We call a model \textit{symmetry-aware} when it explicitly represents and generates space group symmetry, and \textit{symmetry-agnostic} when it places atoms without reference to a space group.

So far, symmetry-aware models have performed worse than symmetry-agnostic ones on standard benchmarks of stability, uniqueness and novelty.
Symmetry restricts how atoms on special Wyckoff positions can move, which can make it harder to find low-energy arrangements.
The typical treatment of coordinates and the lattice can add to this difficulty.
Most models add noise in fractional coordinates, where the physical scale and shape of coordinate perturbations depend on the shape and size of the cell.
Isotropic Cartesian noise avoids this dependence. However, models that use Cartesian diffusion have so far kept the lattice fixed to avoid the difficulty of coupled lattice and coordinate updates \citep{xie2022cdvae}.
High-symmetry crystals with few occupied Wyckoff orbits also share a limited set of structural prototypes.
Generated structures are therefore more likely to repeat one another or match structures already in databases, reducing uniqueness and novelty.

In this work, we introduce Generative Equivariant Orbit Diffusion Engine (GEODE), a symmetry-aware generative model for crystalline materials.
GEODE is, to our knowledge, the first model to combine coordinate and lattice diffusion in Cartesian space.
GEODE first samples a symmetry template, then jointly generates the lattice, atomic coordinates and atom types while preserving the specified symmetry.
We also introduce sampling time template filtering to increase the novelty of generated structures without retraining.
Our main contributions are:
\begin{itemize}
\item We develop a joint diffusion formulation for Cartesian coordinates, lattice parameters and atom types that preserves symmetry throughout the forward and reverse processes, and enables isotropic and consistent noise across different cell dimensions.
\item We introduce a Wyckoff-constrained loss with permutation remapping that restricts coordinate targets to reachable sites and accounts for equivalent atom assignments.
\item We show that GEODE achieves the highest metastable, unique and novel (mSUN) rate among symmetry-aware models, achieving 11.9\% compared with 7.7\% for the next competitor.
\item We show that sampling time template filtering increases mSUN by approximately 6\% without retraining, making GEODE competitive with leading symmetry-agnostic models at negligible additional cost.
\item We demonstrate joint symmetry and property guidance by combining cubic templates with classifier-free guidance to control permittivity while ensuring dielectric isotropy.
\end{itemize}

\section{Related Work}
Most current generative models for inorganic crystals use diffusion to jointly generate atom types, coordinates and the lattice  \citep{xie2022cdvae, jiao2023diffcsp, zeni2025mattergen}.
Recent approaches incorporate space group symmetry by diffusing the asymmetric unit and applying symmetry operations to construct the full crystal.
These models either condition generation on a specified space group \citep{jiao2024space} or first sample the space group and Wyckoff orbits \citep{levy2025symmcd, chang2025sgequidiff}.
WyCryst \citep{zhu2023wycryst, yamazaki2025wycrystplus} and WyckoffDiff \citep{ekstrom2025wyckoffdiff} generate only the prototype, not the full crystal structure, and are therefore not compared here.


\paragraph{SGFM}
\citep{puny2025sgfm} samples a space group and Wyckoff positions from the training distribution, then uses flow matching to generate the lattice, fractional coordinates and atom types.
A group-averaged equivariant vector field preserves Wyckoff constraints without a separate projection step.
GEODE instead generates new symmetry templates and diffuses coordinates in Cartesian space.

\paragraph{SGEquiDiff}
\citep{chang2025sgequidiff} enforces space group symmetry architecturally.
It samples a space group according to its training set frequency, then generates the lattice and Wyckoff orbits with an autoregressive transformer.
An equivariant GNN denoises fractional coordinates within their degrees of freedom.
The lattice and atom types remain fixed during diffusion.
GEODE adapts the Wyckoff transformer for symmetry templates, but jointly diffuses the lattice, coordinates and atom types.

\paragraph{DynaCrys}
\citep{dynacrys} treats the space group as a diffusion variable.
Its symbolic diffusion stage jointly generates the space group, Wyckoff occupations and atom types, using group--subgroup relationships to guide changes in space group.
A second stage built on DiffCSP++ diffuses the lattice and free fractional coordinates.
At the time of writing, we could not find publicly released code or generated structures, so we do not include DynaCrys in our comparisons.

\section{Preliminaries}
\paragraph{Space groups.}
We represent a crystal $C = (\mathbf{A}, \mathbf{X}, \mathbf{L})$ by its atomic species $\mathbf{A} = (a_1, \dots, a_N)$, atomic coordinates $\mathbf{X} = [\mathbf{x}_1, \dots, \mathbf{x}_N] \in \mathbb{R}^{3 \times N}$, and lattice matrix $\mathbf{L} \in \mathbb{R}^{3\times3}$, whose rows are the three vectors spanning the unit cell.
Its space group $\mathcal{G}$ is the set of all rigid motions that map the crystal onto itself~\citep{hahn1983international}.
Each operation $g$ acts as $g \cdot \mathbf{x} = \mathbf{O}\mathbf{x} + \mathbf{t}$, where $\mathbf{O}$ is an orthogonal transformation and $\mathbf{t}$ is a translation.
The crystal is invariant under $g$ if every atom maps to an atom of the same species, so that $g \cdot C = C$ \citep{jiao2024space}.
The orthogonal components $\mathbf{O}$ form the crystal's point group, which constrains the directional dependence of properties such as the dielectric response.
In three dimensions there are 230 possible space groups~\citep{hahn1983international}.

\paragraph{Wyckoff positions.}
Applying all operations in $\mathcal{G}$ to a point $\mathbf{x}$ generates its \textit{orbit}, the set of positions equivalent to $\mathbf{x}$ under symmetry \citep{hahn1983international}.
The operations that leave $\mathbf{x}$ fixed form its \textit{site-symmetry group}, which determines the directions in which the point can move while preserving its site symmetry.
A point fixed only by the identity operation occupies a \textit{general position} and has three positional degrees of freedom.
A point with additional site symmetry occupies a \textit{special position}, where motion is restricted to a plane or line, or the position is fixed entirely.
A \textit{Wyckoff position} groups orbits whose site symmetry groups are related by conjugation within $\mathcal{G}$ \citep{wyckoff1922analytical}.
Each Wyckoff position is identified by a letter and a multiplicity, which gives the number of equivalent sites per orbit in the conventional unit cell \citep{hahn1983international}.
Figure~\ref{fig:overview}b illustrates these concepts for a two-dimensional wallpaper group.

\section{Proposed method: GEODE}
\label{sec:method}

\begin{figure}[t]
    \centering
    \includegraphics[width=0.92\textwidth]{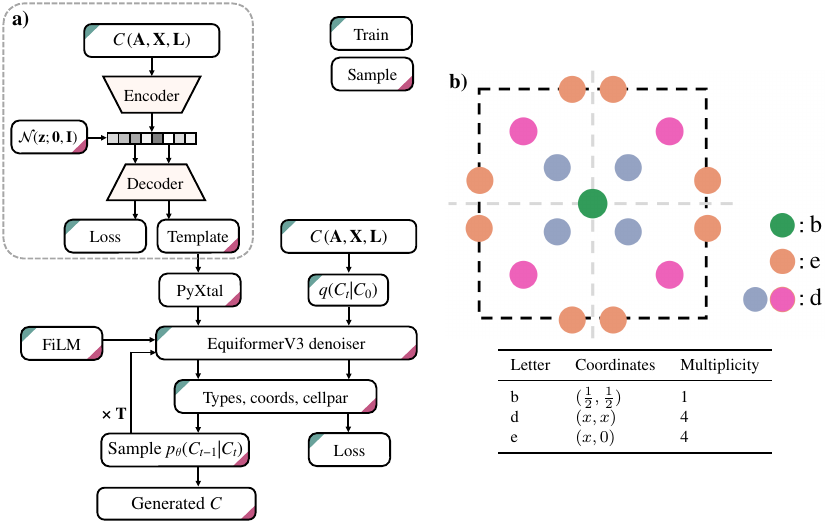}
    \caption{\textbf{(a)} Overview of GEODE.
    The template VAE (dashed box) generates a symmetry template, which PyXtal~\citep{pyxtal} uses to construct an initial structure.
    The diffusion model jointly generates atom types, coordinates and the lattice, preserving symmetry, with optional property conditioning.
    \textbf{(b)} Wyckoff positions of the wallpaper group \textit{p4mm}. Position $b$ is fixed at the cell centre; $d$ and $e$ allow motion along diagonals and horizontal or vertical mirror lines, respectively.}
    \label{fig:overview}
\end{figure}

GEODE generates a crystal in two stages.
First, a Wyckoff template VAE samples a symmetry template $\tau = (G, W)$, consisting of a space group $G$ and a list of Wyckoff orbits $W$.
A diffusion model then generates the full crystal $C = (\mathbf{A}, \mathbf{X}, \mathbf{L})$ conditioned on this template.
The joint distribution over crystals and templates factorises as
\begin{equation}
p(C, \tau) = p_\theta(C \mid \tau)\, p_\psi(\tau),
\qquad
p_\psi(\tau) = \int p_\psi(\tau \mid \mathbf{z})\, \mathcal{N}(\mathbf{z}; \mathbf{0}, \mathbf{I})\, 
\mathrm{d}\mathbf{z},
\end{equation}
where $\mathbf{z}$ is the VAE latent variable.
Marginalising over the discrete set of templates gives the crystal distribution $p(C) = \sum_{\tau} p_\theta(C \mid \tau)\, p_\psi(\tau)$.
During generation we sample $\mathbf{z}$ from the Gaussian prior, decode a template $\tau$ from $\mathbf{z}$, and then generate a crystal from $p_\theta(C \mid \tau)$.
The VAE and diffusion model are trained independently and combined only during generation.
The diffusion model is conditioned only on the template, and both its forward and reverse processes preserve the template's symmetry at every step.
The generated crystal therefore retains all symmetries specified by the sampled template.
Figure~\ref{fig:overview} summarises both stages and Appendix~\ref{si:algorithm} provides the full sampling procedure.

\subsection{Wyckoff template VAE}
\label{sec:vae}
An $E(3)$-equivariant message-passing encoder \citep{batatia2022mace} takes one-hot atom types and Wyckoff site embeddings as input and parameterises a Gaussian posterior over the latent vector $\mathbf{z}$ (Appendix~\ref{si:vae}).
Given $\mathbf{z}$, an MLP decoder predicts a distribution over space groups.
An autoregressive transformer adapted from SGEquiDiff \citep{chang2025sgequidiff} generates the list of $M$ Wyckoff orbits, $W = (w_1, \ldots, w_M)$, conditioned on the latent vector, space group, and preceding entries as
\begin{equation}\label{eq:wyc_factor}
    p_\psi(W \mid G, \mathbf{z}) = \prod_i p_\psi(w_i \mid w_{<i}, G, \mathbf{z}).
\end{equation}
At each step, the transformer predicts Wyckoff letter probabilities subject to space group constraints and a canonical orbit order.
We train the VAE with the objective
\begin{equation}\label{eq:vae_loss}
    \mathcal{L}_{\mathrm{VAE}} = \lambda_{\mathrm{spg}}\,\mathcal{L}_{\mathrm{CE}}
    + \lambda_{\mathrm{wyc}}\,\mathcal{L}_{\mathrm{NLL}}
    + \lambda_{\mathrm{KL}}\, D_{\mathrm{KL}}\!\left(q_\phi(\mathbf{z} \mid C)\,\|\,\mathcal{N}(\mathbf{0}, \mathbf{I})\right),
\end{equation}
where $\mathcal{L}_{\mathrm{CE}}$ is the cross-entropy loss for space group prediction, $\mathcal{L}_{\mathrm{NLL}}$ is the negative log-likelihood of the orbit sequence, and the KL term regularises the posterior towards a standard Gaussian prior.
The coefficients $\lambda_{\mathrm{spg}}$, $\lambda_{\mathrm{wyc}}$ and $\lambda_{\mathrm{KL}}$ weight these terms.

\subsection{Lattice diffusion}
\label{sec:lattice_diffusion}
We diffuse the six cell parameters $\boldsymbol{\ell} = (a, b, c, \alpha, \beta, \gamma)$, the three edge lengths and three angles of the unit cell, rather than the lattice matrix $\mathbf{L}$, since symmetry constraints can be imposed directly on the cell lengths and angles.
We use a variance-preserving process \citep{ho2020denoising} with a cosine schedule \citep{nichol2021improved},
\begin{equation}\label{eq:lattice_forward}
    q(\boldsymbol{\ell}_t \mid \boldsymbol{\ell}_0) = \mathcal{N}\!\left(
    \sqrt{\bar{\alpha}_t}(\boldsymbol{\ell}_0 - \boldsymbol{\mu}) + \boldsymbol{\mu},\;
    (1-\bar{\alpha}_t)\,\mathcal{P}_G\boldsymbol{\Sigma}\mathcal{P}_G^{\top}\right),
\end{equation}
where $\boldsymbol{\Sigma}$ is a diagonal matrix whose entries are the dataset variances of the six cell parameters, scaled down so that sampled cells stay physical (Appendix~\ref{si:lattice_prior}).
At each step, we project the sampled noise onto the subspace allowed by the crystal system (Appendix \ref{si:lattice_constraints}).
The projection $\mathcal{P}_G$ gives equal noise to symmetry-equivalent cell lengths and applies no noise to angles fixed by symmetry.
The prior mean $\boldsymbol{\mu}$ also satisfies these symmetry constraints, and its cell lengths scale with the cube root of the number of atoms.
Appendix~\ref{si:lattice} describes how we construct the prior and apply the projection.

\paragraph{Readout and loss.}
GEODE's EquiformerV3 denoiser is $SE(3)$-equivariant. Its equivariance covers translations and proper rotations, but does not guarantee the required transformation under a mirror, glide or inversion. The projection $\mathcal{R}_G$ imposes the constraints of the other point-group operations (Appendix~\ref{si:equivariance}) on the predicted coordinate updates and we assign the same atom type to all sites in each Wyckoff orbit. 
An equivariant lattice readout predicts a symmetric $3 \times 3$ tensor $\mathbf{S}$ for each structure, with the same transformation properties as a rank-2 stress tensor.
Since the input structure is invariant under the space group of the template and the denoiser is equivariant under rotations, $\mathbf{S}$ is invariant under the rotations of the point group, consistent with Neumann's principle \citep{nye1985physical}.
We use this tensor to deform the noisy lattice as 
\begin{equation}
\hat{\mathbf{L}}_0 = \mathbf{L}_t(\mathbf{I}_3 - \mathbf{S})^{\top}.
\end{equation}
We convert $\hat{\mathbf{L}}_0$ to cell parameters and invert Equation~\ref{eq:lattice_forward} to recover the predicted noise (Equation~\ref{eq:lattice_eps_pred}).
The readout is trained to match the sampled noise, with the residual scaled using $\boldsymbol{\Sigma}$ so that lengths and angles contribute on comparable scales.
Appendix~\ref{si:lattice_prior} provides further details.

\subsection{Atomic coordinate diffusion}
\label{sec:coord_diffusion}

We diffuse atomic coordinates in Cartesian space using a variance-exploding process \citep{song2021scorebased}, which adds noise without drawing positions towards a fixed origin.
Many crystal diffusion models, including DiffCSP and MatterGen \citep{jiao2023diffcsp, zeni2025mattergen}, diffuse fractional coordinates.
However, Cartesian space has the advantage that the noise scale is a real-space distance in \AA{} that is the same for cells of different size (Appendix~\ref{si:cartesian}).
We further adapt the coordinate training target to account for both Wyckoff constraints and permutations of equivalent atoms.

Let $\mathbf{X}_{0\to t}$ denote the clean reference coordinates expressed in the noisy lattice $\mathbf{L}_t$.
We perturb coordinates as $\mathbf{X}_t = \mathbf{X}_{0\to t} + \sigma_t\,\boldsymbol{\epsilon}$ and wrap atoms outside the cell back inside by integer multiples of the lattice vectors.
The noise levels $\sigma_t$ are geometrically spaced, and $\boldsymbol{\epsilon}$ is constructed to preserve symmetry as described below.
We choose the maximum noise level $\sigma_{\max}$ so that the wrapped distribution is uniform over the allowed positional degrees of freedom.

\paragraph{Symmetry-preserving noise.}
We construct correlated noise within each Wyckoff orbit so that the corrupted crystal preserves the template's symmetry.
For a representative atom in orbit $o$, we draw an isotropic vector $\boldsymbol{\eta}_o \sim \mathcal{N}(\mathbf{0}, \mathbf{I}_3)$.
We average this vector over the orthogonal transformations in the atom's site-symmetry group, removing components outside its positional degrees of freedom.
We then transform the projected vector using the symmetry operations and map the vector to the other equivalent atoms.
The resulting vectors form $\boldsymbol{\epsilon}$.
Atoms with no positional degrees of freedom receive no coordinate noise.
Appendix~\ref{si:coord_noise} gives the construction and verifies that it preserves symmetry.

\begin{figure}[t]
    \centering
    \includegraphics[width=\textwidth]{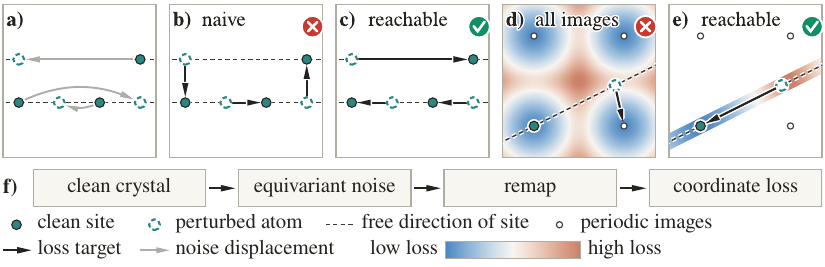}
        \caption{Effect of symmetry constraints on coordinate targets.
        Three atoms of the same species each have one positional degree of freedom, shown by a dashed line.
        Filled circles indicate clean sites and dashed circles corrupted atoms.
        \textbf{(a)} Symmetry-preserving noise moves the two lower atoms past each other.
        \textbf{(b)} Minimising total displacement without Wyckoff constraints assigns atoms to sites they cannot reach within their degrees of freedom.
        \textbf{(c)} Restricting assignments to reachable sites swaps the lower atoms' targets, reducing total displacement compared with restoring the original assignments.
        \textbf{(d)} Wrapped-normal loss around a corrupted atom, with low values in blue and high values in red.
        Including all periodic images gives a gradient that would move the atom off its axis of freedom.
        \textbf{(e)} Using only reachable images keeps the target within the site's free directions.
        \textbf{(f)} Pipeline for obtaining coordinate loss during training.}
    \label{fig:coord_noise}
\end{figure}

\paragraph{Training target.}
We train the coordinate head using a wrapped-normal score target \citep{song2019gradients,zeni2025mattergen}, modified to respect the Wyckoff constraints.
The usual wrapped-normal score combines contributions from periodic copies of the reference position.
For atoms on special positions, some copies cannot be reached by moving along the allowed directions.
Including them can introduce contributions inconsistent with the coordinate constraints (Figure~\ref{fig:coord_noise}d).
We therefore restrict the target to reachable copies, whose displacement to the noisy position lies entirely within the atom's degrees of freedom.

For atom $i$ with displacement $\mathbf{d}_i$ and reachable copies indexed by $\mathcal{I}_i \subseteq \mathbb{Z}^3$, the target is
\begin{equation}\label{eq:coord_target}
    \mathbf{s}^\star_i = -\frac{1}{\sigma_t^2} \sum_{\mathbf{n} \in \mathcal{I}_i} \omega_{\mathbf{n}}\,(\mathbf{d}_i + \mathbf{L}_t^{\top}\mathbf{n}),
    \qquad
    \omega_{\mathbf{n}} \propto \exp\!\left(-\frac{\lVert \mathbf{d}_i + \mathbf{L}_t^{\top}\mathbf{n} \rVert^2}{2\sigma_t^2}\right),
\end{equation}
where the weights $\omega_{\mathbf{n}}$ are normalised over the reachable periodic copies, so that $\sum_{\mathbf{n}\in\mathcal{I}_i} \omega_{\mathbf{n}} = 1$.
The coordinate head predicts the scaled target $\sigma_t\mathbf{s}^\star$ and is trained with $\mathcal{L}_{\mathrm{coord}} = \tfrac{1}{2}\mathbb{E}\lVert \sigma_t\,\mathbf{s}^\star - \hat{\mathbf{s}}_\theta \rVert^2$, where the expectation is over training structures, timesteps and sampled noise.

\paragraph{Permutation remapping.}
Noise can move equivalent atoms closer to one another's reference sites than to their own.
Requiring each atom to return to its original site then makes the target depend on arbitrary atom labelling.
Before computing the coordinate target, we therefore match atoms to reference sites within the same Wyckoff orbit using minimum-cost linear assignment \citep{kuhn1955hungarian}.
The assignment is restricted to reachable sites, so an atom cannot be assigned to a site that requires motion outside its allowed directions (Figure~\ref{fig:coord_noise}b and c).
We use the assigned sites to calculate the displacements and reachable periodic copies in Equation~\ref{eq:coord_target}.
Appendix~\ref{si:remap_proof} analyses this remapping.

\paragraph{Ordering.}
During training and stochastic noise injection at sampling time, we corrupt the lattice first, then the atom types, and finally the coordinates expressed in the corrupted lattice.
For denoising, the coordinate correction is predicted in the frame of the input lattice, so we apply it before updating the lattice and then the atom types.
The sampler therefore undoes the corruption in the reverse of the training order.
The coordinate update is $\mathbf{X}_{t-1} = \mathbf{X}_t + (\sigma_t^2 - \sigma_{t-1}^2)\,\mathcal{R}_G[\hat{\mathbf{s}}_\theta / \sigma_t]$, where $\mathcal{R}_G$ symmetrises the field over the space group to preserve the template's symmetry (Appendix~\ref{si:algorithm}).


\paragraph{Asymmetric unit.}
In order to save computational cost during training and inference, GEODE only computes features for one atom in each orbit and uses the symmetry operations of the crystal to obtain features for other atoms in the same orbit.
The minimal set of inequivalent atoms is called the asymmetric unit (ASU).
Diffusion on the ASU is equivalent to diffusion on the conventional cell.
On MP-20, this triples the generation speed.

\subsection{Atomic type diffusion}
\label{sec:type_diffusion}

We diffuse atom types using an absorbing-state D3PM \citep{austin2021d3pm}.
The process acts on Wyckoff orbits rather than individual atoms, with each orbit sharing a single species variable $a$.
The variable takes one of $K$ element types or the mask state $m$.
During the forward process, an orbit either retains its species or becomes masked.
Once masked, it remains masked for the rest of the forward process. 
Appendix~\ref{si:types} gives the forward process, the loss and the masked-orbit convention.

\section{Experiments}
\label{sec:experiments}

\subsection{Experimental setup}
\label{sec:setup}
We evaluate \emph{de novo} generation on MP-20~\citep{xie2022cdvae, jain2013materials}, a dataset of 45,229 inorganic crystal structures from the Materials Project database with at most 20 atoms in the primitive cell.
We use the 60/20/20 training, validation and test split of \citet{xie2022cdvae}.
We also train on the larger Alex-MP-20 set~\citep{zeni2025mattergen}.
For the generation benchmarks, we evaluate 2,500 structures per model using LeMat-GenBench~\citep{betala2025lematgenbench}.
Structures are classified as stable when their energy above the hull is at or below zero, and metastable when it is greater than zero but no more than 0.1~eV/atom (Appendix~\ref{si:lmb}).
Our primary metric is the mSUN rate, defined as the fraction of metastable, unique and novel structures.
Stable structures are excluded from this metric, because the benchmark treats stable and metastable as disjoint classes and counts the stable, unique and novel structures in a separate SUN rate \citep{betala2025lematgenbench}. 
Several models on the public leaderboard relax structures before submission.
We therefore report results both before and after relaxing our generated structures with the EquiformerV3 potential~\citep{liao2026equiformerv3} trained on OMat24 \citep{barroso2026omat24} and fine-tuned on MPtrj \citep{deng2023chgnet} and sAlex (Appendix~\ref{si:prerelax}).

\begin{figure}[t]
    \centering
    \includegraphics[width=0.8\textwidth]{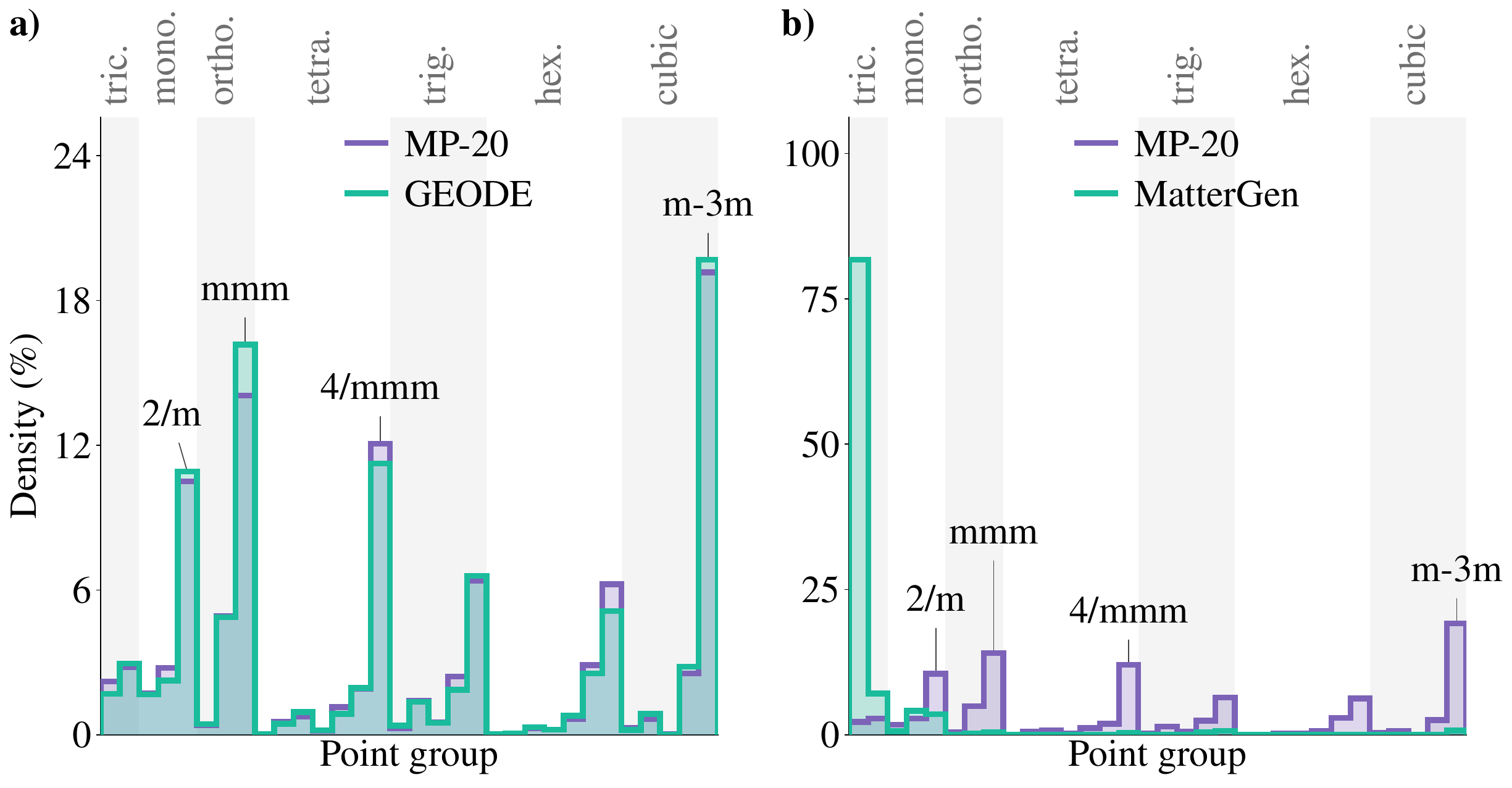}
    \caption{Point-group distribution of \textbf{(a)} GEODE generations and \textbf{(b)} MatterGen structures, compared with the MP-20 training set. Point groups detected on unrelaxed structures at a symmetry tolerance of 0.01~\AA{}. The 32 point groups are ordered by increasing symmetry, with shaded bands indicating the crystal systems. The most common groups are labelled.}
    \label{fig:spg}
\end{figure}

\subsection{Symmetry of the generated crystals}
\label{sec:spg_results}

The template VAE's reconstruction accuracy on the MP-20 validation set is 91.3\% for the space group and 87.8\% for the full Wyckoff occupancy.
Figure~\ref{fig:spg} compares the point-group distribution of structures generated by GEODE and MatterGen with that of the MP-20 training set.
GEODE broadly reproduces the training distribution across the crystal systems.
Cubic symmetry is the most common, while the trivial space group $P1$ accounts for 1.7\% of generations.
At the same symmetry-detection tolerance, 82\% of MatterGen structures are assigned to $P1$ and fewer than 1\% to cubic groups.
On Alex-MP-20, 0.9\% of 10{,}000 GEODE structures are assigned to $P1$, compared with 1.2\% of the training structures, 40.0\% for MatterGen and 99.5\% for Chemeleon~2 (Appendix~\ref{si:alex_pg}).
We also assess whether generated structures preserve their sampled symmetry templates. Of 10,000 generated structures, 98.4\% are assigned exactly the  space group specified by their template.
The remaining 1.6\% are assigned to supergroups of the template space group at the chosen tolerance.

\subsection{De novo generation}
\label{sec:main_results}

Table~\ref{tab:main} compares GEODE with released symmetry-aware generative models, evaluated without structure relaxation.
For most models, we evaluate their released structures.
For SGFM, SGEquiDiff and CrystalFormer we evaluate structures generated using the released MP-20 checkpoints.
GEODE achieves an mSUN rate of 11.92\%, compared with 7.68\% for SGFM, and less than 5\% for the remaining models.
It also has the lowest mean energy above the hull, at 0.196~eV/atom compared with 0.245~eV/atom for SGFM, and ranks among the highest for validity and uniqueness.

We note that symmetry-aware models lag behind leading symmetry-agnostic approaches in mSUN rate (Table~\ref{tab:leaderboard}).
On MP-20, GEODE achieves an mSUN rate of 16.40\% after relaxation, compared with 15.0\% for MatterGen and 22.6\% for Crystalite.
When trained on Alex-MP-20, GEODE reaches an mSUN rate of 29.20\% after relaxation, compared with 62.4\% for Chemeleon~2.
Chemeleon~2 additionally uses reinforcement-learning fine-tuning with rewards for stability and novelty~\citep{park2026chemeleon2}.
After relaxation and template filtering, GEODE achieves the highest SUN rate across all models we tested (Table~\ref{tab:leaderboard}), with a rate of 2.12\% compared with 1.50\% for Crystalite on MP-20, and a rate of 1.32\% compared with 0.96\% for Chemeleon~1 on Alex-MP-20.

\begin{table*}[t]
\centering
\caption{\emph{De novo} generation on MP-20 and Alex-MP-20, evaluated with LeMat-GenBench using 2,500 structures per model.
Metrics follow Section~\ref{sec:setup}.
$E_{\mathrm{hull}}$ is a mean over valid structures.
RMSD measures the root-mean-square atomic displacement after benchmark relaxation.
The WyFormer row uses released structures generated by DiffCSP++ from WyFormer templates.
GEODE rows are in bold; the best value in each column within each dataset is underlined.
$^{*}$ denotes structures relaxed before benchmark evaluation.
Symmetry-agnostic rows are taken from the public LeMat-GenBench leaderboard.
Filtered templates contain 4--6 orbits and occur 1--1,000 times in the MP-20 training set or 11--5,000 times in the Alex-MP-20 training set.
}
\vspace{0.5\baselineskip}
\label{tab:main}
\label{tab:leaderboard}
\label{tab:alex}
\resizebox{\textwidth}{!}{%
\begin{tabular}{l ccc c c c c c}
\toprule
& Valid & Unique & Novel & Metastable & SUN & mSUN & $E_{\mathrm{hull}}$ & RMSD \\
Model & (\%) $\uparrow$ & (\%) $\uparrow$ & (\%) $\uparrow$ & (\%) $\uparrow$ & (\%) $\uparrow$ & (\%) $\uparrow$ & (eV/atom) $\downarrow$ & (\AA) $\downarrow$ \\
\midrule
\multicolumn{9}{c}{\textbf{MP-20}} \\
\midrule
\multicolumn{9}{l}{\textit{Symmetry-agnostic}} \\
\addlinespace[2pt]
Crystalite$^{*}$~\citep{veljkovic2026crystalite} & \underline{97.2} & 95.8 & 53.2 & \underline{51.6} & 1.50 & 22.60 & \underline{0.090} & \underline{0.132} \\
Chemeleon~2~\citep{park2026chemeleon2} & 95.2 & 88.1 & 71.6 & 39.8 & 0.04 & 21.20 & 0.156 & 0.423 \\
MCFlow$^{*}$~\citep{seong2026mcflow} & \underline{97.2} & \underline{96.3} & 52.2 & 49.3 & 0.70 & 18.90 & 0.099 & 0.170 \\
MatterGen$^{*}$~\citep{zeni2025mattergen} & 95.7 & 95.1 & 70.5 & 33.4 & 0.24 & 15.00 & 0.183 & 0.388 \\
DiffCSP~\citep{jiao2023diffcsp} & 95.7 & 94.8 & 66.2 & 29.8 & 0.12 & 8.50 & 0.275 & 0.586 \\
\midrule
\multicolumn{9}{l}{\textit{Symmetry-aware}} \\
\addlinespace[2pt]
SGFM~\citep{puny2025sgfm} & 95.2 & 93.9 & 42.6 & 45.7 & 0.28 & 7.68 & 0.245 & 0.435 \\
DiffCSP++~\citep{jiao2024space} & 95.3 & 95.1 & 62.0 & 26.4 & 0.16 & 4.96 & 0.410 & 0.695 \\
SGEquiDiff~\citep{chang2025sgequidiff} & 91.9 & 91.3 & 58.2 & 22.7 & 0.08 & 2.92 & 0.526 & 0.904 \\
CrystalFormer~\citep{cao2024crystalformer} & 69.7 & 69.1 & 30.8 & 28.0 & 0.00 & 2.56 & 0.669 & 0.646 \\
WyFormer~\citep{kazeev2025wyformer} & 93.7 & 93.3 & 66.2 & 16.5 & 0.08 & 2.40 & 0.489 & 0.772 \\
SymmCD~\citep{levy2025symmcd} & 73.4 & 73.0 & 47.0 & 18.6 & 0.08 & 2.36 & 0.876 & 0.874 \\
\midrule
\textbf{GEODE} & \textbf{95.8} & \textbf{94.9} & \textbf{55.8} & \textbf{41.9} & \textbf{0.62} & \textbf{11.92} & \textbf{0.196} & \textbf{0.419} \\
\textbf{GEODE}$^{*}$ & \textbf{95.6} & \textbf{94.9} & \textbf{54.3} & \textbf{43.7} & \textbf{1.20} & \textbf{16.40} & \textbf{0.135} & \textbf{0.151} \\
\textbf{GEODE-filtered}$^{*}$ & \textbf{96.4} & \textbf{95.3} & \textbf{\underline{74.0}} & \textbf{38.8} & \textbf{\underline{2.12}} & \textbf{\underline{23.80}} & \textbf{0.162} & \textbf{0.203} \\
\midrule
\multicolumn{9}{c}{\textbf{Alex-MP-20}} \\
\midrule
Chemeleon~2~\citep{park2026chemeleon2} & \underline{97.4} & 93.5 & \underline{93.7} & 67.4 & 0.20 & \underline{62.40} & 0.096 & 0.268 \\
MatterGen$^{\dagger}$~\citep{zeni2025mattergen} & 94.3 & 94.3 & 56.3 & 65.3 & 0.80 & 34.48 & \underline{0.079} & 0.139 \\
Chemeleon~1~\citep{park2025chemeleon} & 94.0 & 93.9 & 59.8 & 54.4 & 0.96 & 28.00 & 0.112 & 0.309 \\
CrystaLLM-$\pi$~\citep{bone2025crystallmpi} & 93.4 & 93.2 & 23.9 & \underline{71.3} & 0.32 & 12.48 & 0.094 & 0.191 \\
\midrule
\textbf{GEODE} & \textbf{95.0} & \textbf{\underline{94.9}} & \textbf{54.1} & \textbf{57.7} & \textbf{1.06} & \textbf{25.78} & \textbf{0.102} & \textbf{0.296} \\
\textbf{GEODE}$^{*}$ & \textbf{95.0} & \textbf{\underline{94.9}} & \textbf{52.5} & \textbf{61.8} & \textbf{\underline{1.32}} & \textbf{29.20} & \textbf{0.084} & \textbf{\underline{0.114}} \\
\textbf{GEODE-filtered}$^{*}$ & \textbf{94.8} & \textbf{94.6} & \textbf{64.2} & \textbf{61.7} & \textbf{1.20} & \textbf{36.76} & \textbf{0.086} & \textbf{0.120} \\
\bottomrule
\end{tabular}%
}
\end{table*}

\subsection{Ablations}
\label{sec:ablation}

To identify the source of GEODE's performance gains, we separately test the contributions of Cartesian coordinate diffusion and the modified training target, keeping all other settings fixed (Appendix~\ref{si:frac_ablation}).
Replacing Cartesian with fractional coordinate diffusion lowers mSUN from 11.9\% to 9.3\%, consistently scoring below GEODE across sampling runs (Table~\ref{tab:ablation}).
The decrease mainly reflects stability, with the metastable fraction falling from 41.9\% to 36.5\%, while validity, uniqueness and novelty remain within 3\% of the baseline.
The permutation remapping and the restriction to reachable periodic copies improves mSUN from 11.24 $\pm$ 0.56\% to 11.92 $\pm$ 0.69\% (mean $\pm$ standard deviation over sampling runs).
These changes provide a less noisy training target that respects the coordinate constraints (Appendix~\ref{si:remap_proof}).

\begin{table}[t]
\vspace{-1\baselineskip}
\centering
\small
\setlength{\tabcolsep}{3.2pt}
\renewcommand{\arraystretch}{1.08}
\caption{Impact of coordinate diffusion approach on generation metrics.
In the Cartesian column, checks indicate noise added in Cartesian space and crosses fractional coordinates.
In the Wyckoff loss column, checks indicate permutation remapping with reachable-copy targets and crosses the standard wrapped-normal loss~\citep{zeni2025mattergen}.
Each row is the mean $\pm$ standard deviation over four independent sampling runs containing 2,500 structures.
The best mean in each column is underlined.
}
\vspace{0.5\baselineskip}
\label{tab:ablation}
\begin{tabular}{cc ccccccc}
\toprule
Cartesian & Wyckoff loss & Valid & Unique & Novel & Metastable & SUN & mSUN & $E_{\mathrm{hull}}$ \\
 noise &  and remap & (\%) $\uparrow$ & (\%) $\uparrow$ & (\%) $\uparrow$ & (\%) $\uparrow$ & (\%) $\uparrow$ & (\%) $\uparrow$ & (eV/atom) $\downarrow$ \\
\midrule
\ding{55} & \ding{51} & 95.0\msmall{\pm 0.3} & 94.4\msmall{\pm 0.3} & \underline{58.4\msmall{\pm 0.4}} & 36.5\msmall{\pm 1.4} & 0.50\msmall{\pm 0.12} & 9.28\msmall{\pm 0.78} & 0.249\msmall{\pm 0.007} \\
\ding{51} & \ding{55} & 94.8\msmall{\pm 0.2} & 94.0\msmall{\pm 0.3} & 56.0\msmall{\pm 0.9} & 40.0\msmall{\pm 0.6} & 0.55\msmall{\pm 0.16} & 11.24\msmall{\pm 0.56} & 0.199\msmall{\pm 0.007} \\
\rowcolor{gray!15}
\ding{51} & \ding{51} & \underline{95.8\msmall{\pm 0.3}} & \underline{94.9\msmall{\pm 0.3}} & 55.8\msmall{\pm 0.4} & \underline{41.9\msmall{\pm 0.5}} & \underline{0.62\msmall{\pm 0.15}} & \underline{11.92\msmall{\pm 0.69}} & \underline{0.196\msmall{\pm 0.008}} \\
\bottomrule
\end{tabular}%
\end{table}

\subsection{Steering generation through template filtering}
\label{sec:filter_results}

GEODE separates template generation from crystal generation.
Templates constrain the structural prototypes available to the diffusion model and therefore affect novelty.
We therefore seek to improve novelty by restricting the template distribution $p_\psi(\tau)$ while leaving the diffusion model $p_\theta(C \mid \tau)$ unchanged.
This requires no retraining. Template sampling is about $10^3$ times cheaper than diffusion, so rejecting templates is inexpensive.

Grouping 10,000 generated structures by template reveals relationships between template frequency, orbit count and generation quality (Appendix~\ref{si:template_rationale}).
Templates absent from the training set rarely yield metastable structures, while those occurring more than 1,000 times rarely yield novel ones.
Templates with fewer than four orbits also yield fewer novel structures, while those with more than six yield fewer metastable structures.
These observations motivate retaining templates occurring 1--1,000 times and containing 4--6 orbits.

We generated 2,500 structures per filter combination and evaluated them as in Section~\ref{sec:setup}.
Filtering by frequency raised mSUN from 11.92\% to 13.76\%, by orbit count to 16.40\%, and both together to 18.20\% (Appendix~\ref{si:template_rationale}, Table~\ref{tab:filter}).
With both filters, novelty rose from 55.8\% to 74.6\% while the metastable fraction fell from 41.9\% to 35.1\%.
After relaxation with EquiformerV3, mSUN reached 23.80\%, the highest pre-relaxed MP-20 entry in Table~\ref{tab:main}, compared with 16.40\% unfiltered. We also applied the combined filter to Alex-MP-20, retaining 4--6 orbits but changing the frequency range to 11--5,000 training-set occurrences.
After relaxation, filtering increased mSUN from 29.20\% to 36.76\% and novelty from 52.5\% to 64.2\%, although mSUN remains below that of Chemeleon~2.
Neither experiment required further model training.
Together, these results show that template selection can adjust the balance between novelty and metastability at sampling time.

\subsection{Property steering under a fixed symmetry class}
\label{sec:dielectric_results}

We demonstrate joint symmetry and property guidance by generating cubic crystals with target dielectric constants.
Cubic templates ensure an isotropic dielectric response, while classifier-free guidance controls the permittivity.
We retain cubic templates, about a quarter of those sampled, and fine-tune the MP-20 diffusion model on the AnisoNet dataset \citep{lou_discovery_2025}, training on 5,365 of its 6,706 crystals.
The label is the log of the mean eigenvalue of the electronic dielectric tensor computed with DFT in the Materials Project.
The target is supplied through feature-wise linear modulation (FiLM) layers~\citep{perez2018film}, with a dropout rate of 0.2 during fine-tuning to enable classifier-free guidance~\citep{ho2022cfg}.
The dielectric tensor is not well defined for metallic systems, so we use composition screening to remove structures expected to be metallic and generate structures until about 2,500 per target pass the screen.
This filters up to 55.3\% of generations, with details provided in Appendix~\ref{si:metal_screen}.

\begin{figure}[t]
    \centering
    \includegraphics[width=0.8\textwidth]{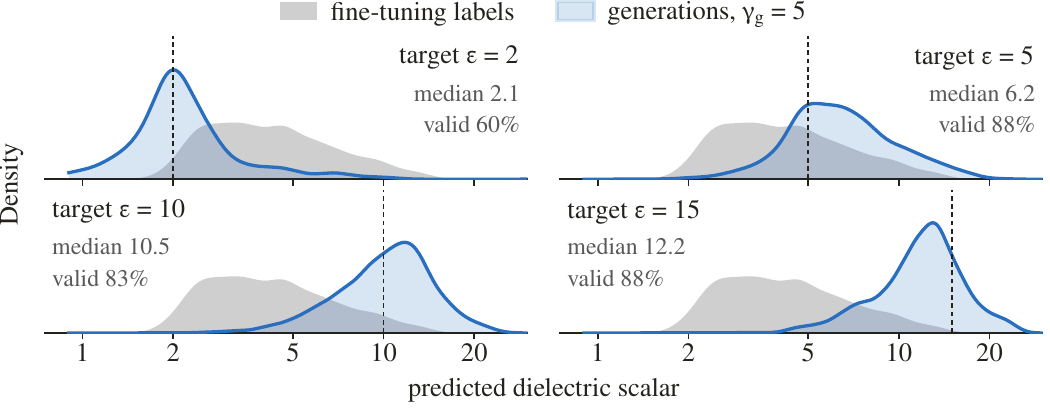}
    \caption{Predicted permittivity of cubic crystals generated with classifier-free guidance at $\gamma_{\mathrm{g}} = 5$.
    Blue distributions show $\sim$2,500 screened structures per target. Grey distributions show the fine-tuning labels.
    Dashed lines indicate the targets.}
    \label{fig:dielectric}
\end{figure}

At guidance strength $\gamma_{\mathrm{g}} = 5$, targets of 2, 5, 10 and 15 give median predicted permittivities of 2.1, 6.2, 10.5 and 12.2, respectively (Figure~\ref{fig:dielectric}).
All structures retain cubic symmetry.
The largest deviation occurs near the upper end of the fine-tuning labels.
As the target increases, compositions shift from fluorides through oxides to nitrides.
Example structures, composition statistics and additional guidance strengths are given in Appendix~\ref{si:dielectric}.
At targets of 5 and above, 83--88\% of screened structures pass the benchmark validity checks, compared with 90\% without property guidance.
At a target of 2, validity falls to 60\%, with many fluorine-rich compositions failing the charge-neutrality check.
Weaker guidance strengths are examined in Appendix~\ref{si:diel_grid}.
AnisoNet supplied both the labels and the evaluation predictions, so these results demonstrate control over the predicted permittivity rather than agreement with DFT.
This control would be challenging for symmetry-agnostic models.

\section{Conclusions and limitations}

We introduced GEODE, a symmetry-aware generative model that samples a symmetry template and then jointly generates the lattice, atomic coordinates and atom types.
GEODE combines Cartesian-space symmetry-preserving denoising with a Wyckoff-constrained loss.
It achieves an mSUN rate of 11.92\% on MP-20 without relaxation, compared with 7.68\% for the next best symmetry-aware model, whilst reproducing the training set point group distribution.
Separating template generation from diffusion allows us to improve generation by selecting templates at sampling time.
Filtering by training-set frequency and orbit count raises mSUN to 23.80\% after relaxation, without further training.
The latter is the highest value on the MP-20 LeMat-GenBench leaderboard as of September 2026.
We also demonstrate joint symmetry and property guidance through generation of cubic materials with target permittivities.
These results show that models with explicit symmetry treatment can support both competitive generation and control over the properties of generated crystals.

The two stages are trained independently, so the template generator receives no feedback on the quality of the resulting crystals.
It may therefore propose templates that the diffusion model has not encountered during training or for which it struggles to generate metastable structures.
Our filter partly addresses this through empirical selection rules.
Future work could investigate training the template generator using the outcomes of crystal generation.
The space group also remains fixed during diffusion, preventing the model from resolving an unsuitable template by changing its symmetry.
An additional direction could be to diffuse the space group and Wyckoff positions jointly with the lattice, coordinates and atom types.
Finally, property guidance was tested for one scalar property within a single crystal system.
Additional tests are required to demonstrate its broader suitability across properties and symmetry classes.

\section*{AI use statement}

We used large language model assistants throughout this work.
They wrote and edited code in our repository, ran and scored experiments under our instructions, produced figures from our results, searched and summarised related literature, drafted parts of the supplementary information from our configuration files and logs, and edited text for readability.
The authors verified all outputs of these assistants, including code, results and references.
The research questions, the model design and the interpretation of the results are the authors' own, and the authors are responsible for the contents of this paper.

\section*{Ethics statement}

This work introduces a generative model for inorganic crystal structures. It
does not involve human subjects, personally identifiable data, or sensitive
data. The training data are the publicly available MP-20 and Alex-MP-20 sets,
and the generated structures are candidate materials whose stability and
properties are predicted, not measured; any use would require experimental
validation.

\section*{Reproducibility statement}

The model and both diffusion processes are specified in Section~\ref{sec:method},
and the sampling procedure is given in full as Algorithm~\ref{alg:sampling} in
Appendix~\ref{si:algorithm}, with all symbols defined in
Appendix~\ref{si:notation}. Training and evaluation details are collected in
Appendices~\ref{si:training} and~\ref{si:eval}.
Code is provided as supplementary material and will be released publicly upon
publication.

\bibliography{refs}
\bibliographystyle{iclr2027_conference}
\clearpage
\appendix
\section{Notation}
\label{si:notation}

\begin{table}[H]
\centering
\caption{Symbols used in the main text and supplementary information.}
\label{tab:notation}
\small
\renewcommand{\arraystretch}{0.95}
\begin{tabular}{@{}lp{0.66\linewidth}@{}}
\toprule
\textbf{Symbol} & \textbf{Meaning} \\
\midrule
\multicolumn{2}{@{}l}{\textit{Crystal and symmetry}} \\
\midrule
$C = (\mathbf{A}, \mathbf{X}, \mathbf{L})$ & Crystal: species, Cartesian coordinates, lattice \\
$N$ & Number of atoms in the cell \\
$\mathbf{f}$ & Fractional coordinates of an atom, $\mathbf{x} = \mathbf{L}^{\top}\mathbf{f}$ \\
$\mathcal{G}$, $g = (\mathbf{O}, \mathbf{t})$ & Space group; one operation, $g \cdot \mathbf{x} = \mathbf{O}\mathbf{x} + \mathbf{t}$ \\
$\tau = (G, W)$ & Template: space group and Wyckoff orbit list \\
$w_i$ & The $i$-th entry of the Wyckoff orbit list $W$, one Wyckoff letter of $G$ \\
$M$ & Number of Wyckoff orbits in a template, $W = (w_1, \ldots, w_M)$ \\
$o$ & One Wyckoff orbit, the set of atoms that one entry $w_i$ generates in the crystal \\
$\mathcal{P}_G$ & Projection onto the cell-parameter subspace allowed by $G$ \\
$\mathcal{R}_G$ & Symmetrisation of a per-atom vector field $\mathbf{v}$ over $G$, Equation~\ref{si:eq:reynolds} \\
\midrule
\multicolumn{2}{@{}l}{\textit{Template VAE}} \\
\midrule
$\mathbf{z}$ & Gaussian latent vector \\
$q_\phi(\mathbf{z} \mid C)$, $p_\psi(\tau \mid \mathbf{z})$ & Encoder; decoder \\
$p_\psi(\tau)$ & Template distribution, the decoder marginalised over the prior on $\mathbf{z}$ \\
$\mathcal{L}_{\mathrm{CE}}$, $\mathcal{L}_{\mathrm{NLL}}$ & Space-group cross-entropy; negative log-likelihood of the orbit list \\
$\lambda_{\mathrm{spg}}, \lambda_{\mathrm{wyc}}, \lambda_{\mathrm{KL}}$ & Weights of the VAE loss terms \\
\midrule
\multicolumn{2}{@{}l}{\textit{Diffusion}} \\
\midrule
$t$, $T$ & Diffusion step; number of steps, $T = 1{,}000$ \\
$p_\theta(C \mid \tau)$ & Diffusion model given a template \\
$\mathbf{A}_t$, $\mathbf{X}_t$, $\mathbf{L}_t$, $\boldsymbol{\ell}_t$ & Species, coordinates, lattice matrix and cell parameters at step $t$ \\
$\boldsymbol{\ell} = (a, b, c, \alpha, \beta, \gamma)$ & Cell parameters \\
$\mathbf{L}(\boldsymbol{\ell})$ & Lattice matrix with cell parameters $\boldsymbol{\ell}$ \\
$\boldsymbol{\mu}$, $\boldsymbol{\Sigma}$ & Cell-parameter prior mean; diagonal covariance before symmetry projection \\
$\bar\ell$, $v_{\mathrm{len}}$ & Mean and variance of training-set cell lengths divided by $N^{1/3}$ \\
$\bar\theta$, $v_{\mathrm{ang}}$ & Mean and variance of training-set free angles \\
$\bar\alpha_t$ & Cumulative product of the lattice diffusion coefficients $\alpha_t$ \\
$\kappa$ & Offset of the cosine schedule \\
$\boldsymbol{\xi}$ & Standard normal vector for the lattice noise \\
$\sigma^{\mathrm{post}}_t$ & Noise scale of the lattice reverse step \\
$\mathbf{S}$, $\hat{\mathbf{L}}_0$, $\hat{\boldsymbol{\ell}}_0$ & Predicted deformation tensor; predicted clean lattice matrix; its cell parameters \\
$\boldsymbol{\epsilon}$ & Symmetry-preserving coordinate noise \\
$\boldsymbol{\epsilon}_{\boldsymbol{\ell}}$, $\hat{\boldsymbol{\epsilon}}_\theta$ & Lattice noise; its prediction \\
$\sigma_t$, $\sigma_{\min}$, $\sigma_{\max}$ & Coordinate noise scale; $\sigma_1 = \sigma_{\min} = 0.01$~\AA, $\sigma_T = \sigma_{\max} = 10$~\AA, and $\sigma_0 = 0$ \\
$\mathcal{N}_{\mathrm{W}}$ & Wrapped normal distribution \\
$\mathbf{X}_{0 \to t}$ & Clean coordinates expressed in the noisy cell \\
$\boldsymbol{\eta}_o$ & Isotropic noise vector of orbit $o$ \\
$\mathbf{d}_i$, $\mathcal{I}_i$ & Displacement of atom $i$; index set of its reachable periodic images \\
$\mathbf{n}$, $\omega_{\mathbf{n}}$ & Index of a periodic image; its wrapped-normal weight \\
$\mathbf{s}^\star$, $\hat{\mathbf{s}}_\theta$ & Coordinate score target; prediction of the scaled target $\sigma_t\mathbf{s}^\star$ \\
$\mathbf{p}$, $\mathbf{t}$, $\bar{\mathbf{t}}$, $\boldsymbol{\delta}$ & Prediction, target, averaged target and $\mathbf{t} - \bar{\mathbf{t}}$ in Appendix~\ref{si:remap_proof} \\
$K$, $m$ & Number of element types, $K = 100$; mask state \\
$a^o_t$ & Species variable of orbit $o$ at step $t$, one of the $K$ elements or $m$ \\
$\beta_t$, $\mathbf{Q}_t$ & Single-step masking probability; single-step transition matrix \\
$\bar\beta_t$ & Probability that an orbit is still unmasked \\
$\mathbf{u}$, $\mathbf{u}^o$, $\boldsymbol{\pi}$ & Type logits; their orbit average; self-conditioning input \\
$N_m$ & Number of atoms in masked orbits \\
$\lambda_{\mathrm{ce}}$ & Weight of the auxiliary cross-entropy \\
$\mathcal{L}_{\mathrm{coord}}$, $\mathcal{L}_{\mathrm{lattice}}$, $\mathcal{L}_{\mathrm{type}}$ & Coordinate, lattice and type losses \\
\midrule
\multicolumn{2}{@{}l}{\textit{Guidance and evaluation}} \\
\midrule
$\gamma_{\mathrm{g}}$, $\varepsilon$ & Guidance strength; target permittivity \\
$E_{\mathrm{hull}}$ & Energy above the convex hull \\
\bottomrule
\end{tabular}
\end{table}

\section{Sampling algorithm}
\label{si:algorithm}

Algorithm~\ref{alg:sampling} gives the full sampling procedure.
Line 1 decodes a template from a latent vector sampled from the prior.
Lines 2 and 3 sample a cell from the lattice prior, place the orbits in it with PyXtal and add symmetry-preserving noise at the largest noise scale.
Line 4 masks every orbit and sets the self-conditioning input $\boldsymbol{\pi}$ to zero.
Lines 5 to 18 are the diffusion loop.
At each step the denoiser predicts the scaled coordinate target $\hat{\mathbf{s}}_\theta$, a cell deformation $\mathbf{S}$ and per-atom type logits $\mathbf{u}$ (line 6).
Line 6 also sets $\boldsymbol{\pi}$ to the softmax of $\mathbf{u}$ for the next step (Appendix~\ref{si:training}).
Line 7 moves the atoms by one variance-exploding reverse step.
The step is averaged over the space group by $\mathcal{R}_G$, so the displacements of symmetry-equivalent atoms are related by the space-group operations and every atom remains on its Wyckoff site.
Line 8 computes the predicted clean cell from the deformation and then the predicted lattice noise from that cell.
Line 9 takes the DDPM mean step \citep{ho2020denoising} on the cell parameters, projected onto the subspace allowed by the space group.
With the linear masking schedule $\bar\beta_t = 1 - t/T$, lines 10 to 13 unmask each masked orbit with probability $1/t$, so every orbit is unmasked by $t = 1$.
The element of each newly unmasked orbit is sampled from the orbit-averaged logits of the 76 allowed elements (Appendix~\ref{si:types}) and assigned to all atoms of the orbit.
Unmasked orbits keep their element, and each atom in $\mathbf{A}_t$ takes the element of its orbit.
Lines 14 to 17 add newly sampled noise at every step except the last.
The lattice is noised before the coordinates, as in training.
The coordinate noise of line 16 is built in the cell updated by line 15.
The schedules end at $\sigma_0 = 0$ and $\bar\alpha_0 = 1$, and the lattice step uses $\alpha_t = \bar\alpha_t / \bar\alpha_{t-1}$, $\sigma^{\mathrm{post}}_t = \sqrt{(1 - \alpha_t)(1 - \bar\alpha_{t-1})/(1 - \bar\alpha_t)}$ and $\boldsymbol{\xi} \sim \mathcal{N}(\mathbf{0}, \mathbf{I}_6)$.
The cell matrix with parameters $\boldsymbol{\ell}$ is written $\mathbf{L}(\boldsymbol{\ell})$.
Whenever the cell changes, the fractional coordinates of the atoms are held fixed.
The operator $\mathcal{R}_G$ of line 7 averages a per-atom vector field $\mathbf{v}$ over the finite set of operations of $G$ modulo the conventional-cell translations,
\begin{equation}\label{si:eq:reynolds}
    (\mathcal{R}_G \mathbf{v})_j = \frac{1}{|G|} \sum_{g \in G} \mathbf{O}_g\, \mathbf{v}_{g^{-1} j},
\end{equation}
where $g^{-1} j$ is the atom that $g$ maps onto site $j$, with positions identified modulo conventional-cell translations.
Under classifier-free guidance towards a property target, the denoiser output is the unconditional output plus $\gamma_{\mathrm{g}}$ times the difference between the conditional and unconditional outputs, so $\gamma_{\mathrm{g}} = 1$ recovers the conditional model.

\begin{algorithm}[H]
\caption{GEODE sampling.}
\label{alg:sampling}
\begin{algorithmic}[1]
    \State $\mathbf{z} \sim \mathcal{N}(\mathbf{0}, \mathbf{I})$;\quad $\tau = (G, W) \gets \text{decoder}(\mathbf{z})$
    \State $\boldsymbol{\ell}_T \gets \boldsymbol{\mu} + \mathcal{P}_G[\boldsymbol{\Sigma}^{1/2}\boldsymbol{\xi}]$;\quad $\mathbf{L}_T \gets \mathbf{L}(\boldsymbol{\ell}_T)$
    \State $\mathbf{X}_T \gets \text{PyXtal}(\tau, \mathbf{L}_T) + \sigma_T\,\boldsymbol{\epsilon}$
    \State $a^o_T \gets m$ for all $o \in W$;\quad $\boldsymbol{\pi} \gets \mathbf{0}$
    \For{$t = T, \dots, 1$}
        \State $(\hat{\mathbf{s}}_\theta, \mathbf{S}, \mathbf{u}) \gets \text{denoiser}(\mathbf{X}_t, \mathbf{L}_t, \mathbf{A}_t, \boldsymbol{\pi}, t \mid \tau)$;\quad $\boldsymbol{\pi} \gets \mathrm{softmax}(\mathbf{u})$
        \State $\mathbf{X}_{t-1} \gets \mathbf{X}_t + (\sigma_t^2 - \sigma_{t-1}^2)\, \mathcal{R}_G[\hat{\mathbf{s}}_\theta / \sigma_t]$
        \State $\hat{\mathbf{L}}_0 \gets \mathbf{L}_t(\mathbf{I}_3 - \mathbf{S})^\top$;\quad $\hat{\boldsymbol{\epsilon}}_\theta \gets$ Eq.~\ref{eq:lattice_eps_pred}
        \State $\boldsymbol{\ell}_{t-1} \gets \boldsymbol{\mu} + \mathcal{P}_G\!\left[\tfrac{1}{\sqrt{\alpha_t}}\big((\boldsymbol{\ell}_t - \boldsymbol{\mu}) - (1 - \alpha_t)(1 - \bar\alpha_t)^{-1/2}\, \hat{\boldsymbol{\epsilon}}_\theta\big)\right]$;\quad $\mathbf{L}_{t-1} \gets \mathbf{L}(\boldsymbol{\ell}_{t-1})$
        \ForAll{$o \in W$ with $a^o_t = m$}
            \State $\mathbf{u}^o \gets \frac{1}{|o|} \sum_{i \in o} \mathbf{u}_i$
            \State $a^o_{t-1} \sim \mathrm{softmax}(\mathbf{u}^o)$ with probability $(\bar\beta_{t-1} - \bar\beta_t)/(1 - \bar\beta_t) = 1/t$, else $a^o_{t-1} \gets m$
        \EndFor
        \If{$t > 1$}
            \State $\boldsymbol{\ell}_{t-1} \gets \boldsymbol{\ell}_{t-1} + \sigma^{\mathrm{post}}_t\, \mathcal{P}_G[\boldsymbol{\Sigma}^{1/2}\boldsymbol{\xi}]$;\quad $\mathbf{L}_{t-1} \gets \mathbf{L}(\boldsymbol{\ell}_{t-1})$
            \State $\mathbf{X}_{t-1} \gets \mathbf{X}_{t-1} + \sqrt{\sigma_t^2 - \sigma_{t-1}^2}\,\boldsymbol{\epsilon}$
        \EndIf
    \EndFor
    \State \Return $(\mathbf{A}_0, \mathbf{X}_0, \mathbf{L}_0)$
\end{algorithmic}
\end{algorithm}

\section{Wyckoff template VAE details}
\label{si:vae}

\paragraph{Encoder.}
The encoder is a two-layer MACE \citep{batatia2022mace}, a message-passing graph neural network.
Each node represents an atom of the conventional cell and is connected by edges to its neighbours within 7~\AA{}, including periodic images (Appendix~\ref{si:training}).
Each layer of the network performs one round of message passing over this graph.
The input features of each atom are a one-hot encoding of its element and the 231-dimensional embedding of its Wyckoff position from SGEquiDiff \citep{chang2025sgequidiff}.
The encoder receives the space group only through this embedding, which also encodes the site-symmetry group, the multiplicity and the number of free parameters.
Features read out from every atom after each layer are summed into a 256-dimensional crystal vector that is invariant under rotation, translation and permutation of the atoms.
Two linear layers map this vector to the mean and log-variance of $q_\phi(\mathbf{z} \mid C)$, a Gaussian over the 256-dimensional latent vector.
An MLP with two hidden layers of width 512 maps $\mathbf{z}$ to logits over the 230 space groups.

\paragraph{Wyckoff transformer.}
The Wyckoff transformer of Equation~\ref{eq:wyc_factor} follows SGEquiDiff, but is conditioned on $\mathbf{z}$ instead of the lattice and emits Wyckoff letters only, without atom types.
Each token is the SGEquiDiff embedding of one orbit's Wyckoff position, concatenated with the output of a two-layer MLP applied to $\mathbf{z}$.
The sequence passes through one causal transformer layer of width 256 with four attention heads.
The distribution over the next letter is the attention of the current token over the letters of the space group and a stop token, so the space group fixes the vocabulary.
One mask keeps the letters in non-decreasing alphabetical order, the canonical order of Section~\ref{sec:vae}.
A second mask allows a letter to repeat only if its Wyckoff position has a free parameter, since two orbits of a Wyckoff position without free parameters would occupy the same sites.
The sequence cannot stop at the first step and ends after at most 20 orbits.
The loss $\mathcal{L}_{\mathrm{NLL}}$ is the negative log-likelihood of each letter and of the stop token given the true preceding letters, averaged over tokens.

\paragraph{Decoding.}
A template is decoded from $\mathbf{z}$ by taking the most probable space group from the space-group MLP and generating the letters with greedy decoding under the masks.
A template that PyXtal cannot realise, or whose realised structure is not detected in the requested space group, is discarded before diffusion.

\paragraph{Training.}
We train one template VAE on MP-20 and one on Alex-MP-20, each with 2.68~million parameters.
The loss weights in Equation~\ref{eq:vae_loss} are $\lambda_{\mathrm{spg}} = 1$, $\lambda_{\mathrm{wyc}} = 10$ and $\lambda_{\mathrm{KL}} = 0.3$.
Training uses Adam with a learning rate of $10^{-3}$, batches of 128 crystals and gradient clipping at norm 1.
The learning rate is reduced by a factor of 0.6 when the validation loss stops improving.
Sampling uses an exponential moving average of the parameters with decay 0.99, taken from the checkpoint with the lowest validation loss.
On the Alex-MP-20 validation set, the reconstruction accuracy is 92.6\% for the space group and 93.4\% for the full Wyckoff occupancy.

\section{Lattice diffusion details}
\label{si:lattice}

\subsection{Cell-parameter prior}
\label{si:lattice_prior}
The prior $\mathcal{N}(\boldsymbol{\mu}, \boldsymbol{\Sigma})$ is built from MP-20 training-set statistics and depends on the crystal system and on the number of atoms $N$ in the cell.
For the lengths we divide every training-set cell length by $N^{1/3}$ before taking the mean $\bar{\ell}$ and the variance $v_{\mathrm{len}}$, and set $\mu_a = \mu_b = \mu_c = \bar{\ell}\, N^{1/3}$.
The cube of this quantity is proportional to the volume per atom, so the prior sets the volume per atom, rather than the absolute cell size, to the dataset average.
Angles fixed by the crystal system are set to exactly $90^\circ$ or $120^\circ$ independently of the prior, so the angle statistics need to describe only the free angles.
We discard training entries within $0.1^\circ$ of $90^\circ$ or $120^\circ$ and take the mean $\bar\theta$ and variance $v_{\mathrm{ang}}$ of the rest.
The prior mean is $\bar\theta$ for each free angle and the value set by the crystal system for each fixed angle.
The mean therefore lies in the affine set of cell parameters compatible with the crystal system, so projecting a perturbed mean back onto that set and subtracting the mean is the same as applying $\mathcal{P}_G$ to the perturbation alone.
The covariance $\boldsymbol{\Sigma}$ is diagonal and shared across crystal systems, with $v_{\mathrm{len}}$ on the three length entries and $v_{\mathrm{ang}}$ on the three angle entries, both divided by 64.
Without this factor, samples from the prior occasionally have near-zero or negative lengths, which make atoms overlap, or extreme angles, which produce degenerate cells.
On the MP-20 training set the constants are $\bar{\ell} = 2.94$~\AA{}, $\bar\theta = 99.7^\circ$, $v_{\mathrm{len}} = 2.52$~\AA$^2$ and $v_{\mathrm{ang}} = 297$~deg$^2$.
The prior mean of a cell with $N$ atoms is therefore $a = b = c = 2.94\, N^{1/3}$~\AA{}, which is 5.9~\AA{} at $N = 8$ and 8.0~\AA{} at $N = 20$, with angles $(90, 90, 90)^\circ$ for cubic, tetragonal and orthorhombic cells, $(90, 99.7, 90)^\circ$ for monoclinic cells, $(90, 90, 120)^\circ$ for trigonal and hexagonal cells and $(99.7, 99.7, 99.7)^\circ$ for triclinic cells.

The lattice loss is computed on the noise estimate implied by the prediction of the lattice readout. Converting the predicted cell $\hat{\mathbf{L}}_0$ to cell parameters $\hat{\boldsymbol{\ell}}_0$ and inverting the forward process gives
\begin{align}
    \hat{\boldsymbol{\epsilon}}_\theta &= \frac{(\boldsymbol{\ell}_t - \boldsymbol{\mu})- \sqrt{\bar{\alpha}_t}\,(\hat{\boldsymbol{\ell}}_0 - \boldsymbol{\mu})}{\sqrt{1-\bar{\alpha}_t}}, \label{eq:lattice_eps_pred} \\
    \mathcal{L}_{\mathrm{lattice}} &= \left\lVert \boldsymbol{\Sigma}^{-1/2} (\hat{\boldsymbol{\epsilon}}_\theta - \boldsymbol{\epsilon}_{\boldsymbol{\ell}}) \right\rVert^2, \label{eq:lattice_loss}
\end{align}
 where $\boldsymbol{\Sigma}^{-1/2}$ normalises each component of the residual by the prior noise scale of its cell parameter.

\subsection{Constraints per crystal system}
\label{si:lattice_constraints}
The projection $\mathcal{P}_G$ acts on a six-component vector of cell-parameter noise.
It assigns the same noise component to all cell lengths that the crystal system constrains to be equal, and zero noise to all angles that the crystal system fixes.
The perturbed cell parameters therefore always satisfy the constraints of that crystal system.
Table~\ref{tab:lattice_constraints} lists the free parameters in the conventional setting used throughout.
Trigonal space groups, including the rhombohedral ones, are described in the hexagonal setting.

\begin{table}[H]
\centering
\caption{Cell-parameter constraints by crystal system, conventional cell.}
\label{tab:lattice_constraints}
\small
\begin{tabular}{@{}lll@{}}
\toprule
Crystal system & Free parameters & Constraints \\
\midrule
Triclinic & $a, b, c, \alpha, \beta, \gamma$ & none \\
Monoclinic & $a, b, c, \beta$ & $\alpha = \gamma = 90^\circ$ \\
Orthorhombic & $a, b, c$ & $\alpha = \beta = \gamma = 90^\circ$ \\
Tetragonal & $a, c$ & $b = a$; $\alpha = \beta = \gamma = 90^\circ$ \\
Trigonal and hexagonal & $a, c$ & $b = a$; $\alpha = \beta = 90^\circ$, $\gamma = 120^\circ$ \\
Cubic & $a$ & $b = c = a$; $\alpha = \beta = \gamma = 90^\circ$ \\
\bottomrule
\end{tabular}
\end{table}

\subsection{Forward process and training target}
\label{si:lattice_ddpm}
A forward step samples $\boldsymbol{\xi} \sim \mathcal{N}(\mathbf{0}, \mathbf{I}_6)$, scales it by $\boldsymbol{\Sigma}^{1/2}$ and projects it with $\mathcal{P}_G$.
Cell lengths constrained to be equal are all assigned the noise component sampled for the first of them rather than an average, which keeps the variance of each free length at its prior value.
The noise component of every angle fixed by the crystal system is set to zero.
In a cubic cell the vector $(\epsilon_a, \epsilon_b, \epsilon_c, \epsilon_\alpha, \epsilon_\beta, \epsilon_\gamma)$ becomes $(\epsilon_a, \epsilon_a, \epsilon_a, 0, 0, 0)$.
The projection leaves unchanged any perturbation that already satisfies the constraints of the crystal system.
The added noise therefore has covariance $\mathcal{P}_G \boldsymbol{\Sigma} \mathcal{P}_G^{\top}$ and perturbs the cell parameters only within the subspace allowed by the crystal system.
The noise on the free parameters is still Gaussian with its prior variance, so the process on those parameters is an ordinary DDPM in the variable $\boldsymbol{\Sigma}^{-1/2}(\boldsymbol{\ell} - \boldsymbol{\mu})$ and the standard reverse step applies unchanged.
Both the noise and the clean cell parameters are whitened, with the mean $\boldsymbol{\mu}$ removed, so the variance-preserving step mixes two zero-mean variables.
The mixing coefficients $\sqrt{\bar\alpha_t}$ and $\sqrt{1 - \bar\alpha_t}$ do not sum to one, so without removing $\boldsymbol{\mu}$ the mean would drift.
Per sample, Equation~\ref{eq:lattice_forward} reads
\begin{equation}\label{si:eq:lattice_forward}
    \boldsymbol{\ell}_t = \sqrt{\bar\alpha_t}\,(\boldsymbol{\ell}_0 - \boldsymbol{\mu}) + \sqrt{1 - \bar\alpha_t}\,\boldsymbol{\epsilon}_{\boldsymbol{\ell}} + \boldsymbol{\mu},
    \qquad \boldsymbol{\epsilon}_{\boldsymbol{\ell}} = \mathcal{P}_G[\boldsymbol{\Sigma}^{1/2}\boldsymbol{\xi}],
\end{equation}
and $\boldsymbol{\epsilon}_{\boldsymbol{\ell}}$ is the training target of Equation~\ref{eq:lattice_loss}.
The noisy parameters are converted to a cell matrix and assigned to the structure with fractional coordinates preserved.
The loss is formed in cell-parameter space rather than on the tensor $\mathbf{S}$ so that the quantity being regressed is the same noise the forward process added.
The cosine schedule \citep{nichol2021improved} uses the offset $\kappa = 0.008$ with $1 - \alpha_t$ clipped at $0.999$, so $\bar\alpha_t$ falls from $1 - 4 \times 10^{-5}$ at $t = 1$ to $2 \times 10^{-9}$ at $t = T$.

\section{Atomic coordinate diffusion details}
\label{si:coord}

\subsection{Cartesian versus fractional coordinates}
\label{si:cartesian}
Fractional coordinates $\mathbf{f}$ give each atomic position as fractions of the three cell vectors, and the Cartesian position is $\mathbf{x} = \mathbf{L}^{\top}\mathbf{f}$, since the rows of $\mathbf{L}$ are the cell vectors.
We chose to diffuse Cartesian coordinates, but other models in the field, such as MatterGen \citep{zeni2025mattergen}, diffuse fractional coordinates.
Here we discuss the implications of both approaches.
Fractional coordinates range from 0 to 1 along each cell vector and wrap around at the cell edge.
Noise of scale $\sigma_t$ added to them and wrapped back into the cell therefore gives a distribution that is independent of the cell,
\begin{equation}\label{si:eq:frac_kernel}
    q(\mathbf{f}_t \mid \mathbf{f}_0) = \mathcal{N}_{\mathrm{W}}\big(\mathbf{f}_t;\, \mathbf{f}_0,\, \sigma_t^2\mathbf{I}\big).
\end{equation}
A noisy sample at any step can then be sampled directly from the clean crystal, and the forward process factorises into independent processes for the coordinates, the lattice and the types.

Cartesian coordinates lose this factorisation once the cell is noised as well.
An atom at a fixed fractional position moves in Cartesian space whenever the cell changes, so a Cartesian step combines the noise added to the atom with the displacement caused by the change of cell.
If the noise is added step by step, these contributions accumulate, and the noisy coordinates at step $t$ become
\begin{equation}\label{si:eq:cart_path_kernel}
    q(\mathbf{x}_t \mid \mathbf{x}_0, \mathbf{L}_{1:t}) = \mathcal{N}_{\mathrm{W}}\Big(\mathbf{x}_t;\, \mathbf{L}_t^{\top}\mathbf{L}_0^{-\top}\mathbf{x}_0,\, \mathbf{L}_t^{\top}\Big(\sum_{t'=1}^{t} (\sigma_{t'}^2 - \sigma_{t'-1}^2)\,\mathbf{L}_{t'}^{-\top}\mathbf{L}_{t'}^{-1}\Big)\mathbf{L}_t\Big),
\end{equation}
wrapped in the cell $\mathbf{L}_t$.
This kernel depends on the whole corruption path of the cell.
Sampling one training example therefore means simulating that path.
\citet{zeni2025mattergen} also report that the sum is numerically unstable over long paths.

We diffuse Cartesian coordinates so that the noise scale is a real-space distance in \AA ngstr\"om.
A fractional displacement of fixed size is a different physical distance in every cell, and a different one along each axis of a non-cubic cell.
A Cartesian $\sigma_t$ is the same length in every crystal.
We avoid simulating the path by defining the noisy coordinates directly instead of deriving them from a step-by-step process.
For each training example we sample one step $t$, sample the noisy cell $\mathbf{L}_t$ directly from the clean cell, express the clean atoms in it with the fractional coordinates unchanged, and add one Cartesian displacement of scale $\sigma_t$ in that cell,
\begin{equation}\label{si:eq:cart_defined_kernel}
    q(\mathbf{x}_t \mid \mathbf{x}_0, \mathbf{L}_t) = \mathcal{N}_{\mathrm{W}}\big(\mathbf{x}_t;\, \mathbf{L}_t^{\top}\mathbf{L}_0^{-\top}\mathbf{x}_0,\, \sigma_t^2\mathbf{I}\big),
\end{equation}
wrapped in $\mathbf{L}_t$.
The noise is restricted to the free directions of each site (Appendix~\ref{si:coord_noise}).
Denoising score matching needs only this kernel, so no earlier step is simulated.

Our forward process does not factorise into independent coordinate and lattice processes.
This drawback was previously identified by \citet{zeni2025mattergen}.
The reverse steps of Algorithm~\ref{alg:sampling} move the atoms and then the cell.
They are not guaranteed to be the exact reverse of Equation~\ref{si:eq:cart_defined_kernel}.
We accept both drawbacks.
The denoiser is trained to reverse the coupled corruption, and sampling corrects the coordinates in the current cell before the cell changes.
The reduced prior covariance of Appendix~\ref{si:lattice_prior} keeps cells close to the dataset average, which is sufficient for stable coordinate training.
With everything else unchanged, diffusing fractional coordinates instead lowers mSUN from 11.9\% to 9.3\%, mainly through a lower metastable fraction (Appendix~\ref{si:frac_ablation}).

\subsection{Noise schedule and prior}
\label{si:coord_schedule}
The noise scale $\sigma_t$ is geometric in $t$ from $\sigma_{\min} = 0.01$~\AA{} to $\sigma_{\max} = 10$~\AA{} over $T = 1{,}000$ steps.
Whenever the cell changes, the atoms keep their fractional coordinates, both in training and sampling.
The lattice process therefore never moves an atom relative to its Wyckoff site.
Positions are wrapped into the cell after every step, so a Gaussian displacement whose scale exceeds the cell dimensions smears coordinates into a uniform distribution along their degrees of freedom.

\begin{figure}[H]
\centering
\includegraphics[width=0.6\textwidth]{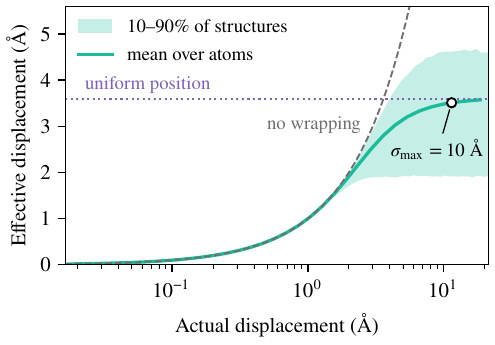}
\caption{Effective displacement of an atom after wrapping into the cell, against its actual displacement before wrapping, under the symmetry-preserving coordinate noise. The atoms are those with at least one free direction in 1,000 MP-20 training structures. The line is the mean over atoms and the band spans the 10th to 90th percentiles across structures. The dashed line is the displacement without wrapping, the dotted line the mean distance of a uniformly placed atom from its site, and the marker $\sigma_{\max} = 10$~\AA{}.}
\label{fig:eff_disp}
\end{figure}

Figure~\ref{fig:eff_disp} shows that $\sigma_{\max} = 10$~\AA{} is large enough for MP-20 structures.
It compares the actual and the effective displacement of each atom over 1,000 MP-20 training structures.
The actual displacement is measured before wrapping into the cell, and the effective displacement after wrapping.
The two agree up to about 1.5~\AA{}.
Beyond that the effective displacement levels off at 3.6~\AA{}, the mean distance of a uniformly placed atom from its site.
The value of $\sigma_{\max}$ is chosen to leave a plateau at the higher noise levels, to accommodate structures with larger unit cells or structures with many positional degrees of freedom.

Sampling starts from the symmetry template. PyXtal \citep{pyxtal} places each orbit at a position sampled uniformly along its free directions, and the symmetry-preserving noise at $\sigma_{\max}$ is then added.

\subsection{Symmetry-preserving noise}
\label{si:coord_noise}
The noise of Section~\ref{sec:coord_diffusion} moves each atom only along the directions its Wyckoff site allows.
We sample one random vector per orbit and average it over the site-symmetry group of the orbit's representative atom.
The space-group operations then transform the result and map it to the other atoms of the orbit.
Together these two steps are one application of the operator $\mathcal{R}_G$ of Equation~\ref{si:eq:reynolds}.
The displacement of each atom is Gaussian with variance $\sigma_t^2$ along each of its free directions and zero along the others.
The displacements of all atoms of an orbit are derived from one random vector, so the noisy crystal preserves the template's space group.
The noise scale is the same along every free direction of every site, so one wrapped-normal target, Equation~\ref{eq:coord_target}, applies to every site.
Each atom's target is the score of the orbit's shared displacement, up to a factor equal to the number of atoms in the orbit.
The loss averages over atoms, so it weights each orbit by its number of atoms.

\subsection{Permutation remapping}
\label{si:remap_proof}

\paragraph{Assignment.}
After the coordinate noise is added in training, remapping matches each perturbed atom to a site of its orbit.
It selects the matching with the minimum total displacement, with each distance measured to the nearest periodic image.
The atoms of an orbit are identical, so any site of the orbit is a valid target.
Unreachable pairs are excluded, such as ones that are not on an atom's axis or plane of freedom, and the Hungarian algorithm \citep{kuhn1955hungarian} finds the minimum-cost matching among the remaining pairs within each orbit.
Remapping is used only to form the training target, not in sampling.

\paragraph{Remapping.}
Two atoms of the same orbit can swap places during the perturbation, and the network cannot tell them apart.
For one atom, write $\mathbf{p}$ for the prediction of the coordinate head and $\mathbf{t}$ for the target computed from the site the atom actually came from.
Write $\bar{\mathbf{t}}$ for the average of $\mathbf{t}$ over the sites of the orbit the atom could have come from, each weighted by the probability that the noise displaced the atom from that site.
The average target $\bar{\mathbf{t}}$ depends only on the noisy crystal, which is the only input to the network.
Write $\boldsymbol{\delta} = \mathbf{t} - \bar{\mathbf{t}}$ for the part of the target the noisy crystal does not determine.
Substituting $\mathbf{t} = \bar{\mathbf{t}} + \boldsymbol{\delta}$ and expanding the square gives
\begin{align}
    \tfrac{1}{2}\lVert \mathbf{p} - \mathbf{t} \rVert^2
    &= \tfrac{1}{2}\lVert (\mathbf{p} - \bar{\mathbf{t}}) - \boldsymbol{\delta} \rVert^2 \nonumber \\
    &= \tfrac{1}{2}\lVert \mathbf{p} - \bar{\mathbf{t}} \rVert^2
       - (\mathbf{p} - \bar{\mathbf{t}})\cdot\boldsymbol{\delta}
       + \tfrac{1}{2}\lVert \boldsymbol{\delta} \rVert^2. \label{si:eq:remap_split}
\end{align}
The first term is the part the network can learn, and the minimum is at $\mathbf{p} = \bar{\mathbf{t}}$.
The second term averages to zero over many samples.
For a fixed model and a given noisy crystal, the model produces the same $\mathbf{p}$ every time and $\bar{\mathbf{t}}$ is fixed, so $\mathbf{p} - \bar{\mathbf{t}}$ is constant.
The mean of $\mathbf{t}$ over many samples approaches $\bar{\mathbf{t}}$ by definition, thus the mean of $\delta$ approaches zero, and the second term averages to zero.
However, the second term will not be zero for individual noise samples, thus including it will add unlearnable noise to the loss.
The last term contains no network output, so it only adds a floor to the loss.
Changing the loss weight cannot compensate for the second and the third term, since coordinate loss is competing for gradients with type and lattice loss, and these two terms contribute nothing useful to the gradients.
The remapping uses the target of the nearest site, which equals $\bar{\mathbf{t}}$ when one site clearly dominates.
The remapped loss then keeps only the first term, and otherwise approximates it with a small bias.

\subsection{Fractional-coordinate ablation}
\label{si:frac_ablation}
The fractional variant of Table~\ref{tab:ablation} changes only the space in which the coordinates are diffused and keeps everything else as in GEODE.
The noise follows DiffCSP++ \citep{jiao2024space}.
For each Wyckoff orbit, one Gaussian vector is sampled in fractional coordinates, restricted to the allowed directions of the site and mapped to the other atoms of the orbit by the space-group operations.
The network still predicts a Cartesian vector, which is converted to fractional coordinates with the inverse of the noisy cell before the loss, as in MatterGen \citep{zeni2025mattergen}.
The noise scale ranges from 0.005, the DiffCSP++ value, to $\sqrt{5}/2$ in fractional units.
The upper end is higher than the DiffCSP++ value of 0.5, since the noise had to be adapted to GEODE's framework.
Our sampler starts from atoms distributed uniformly along their free directions. A free line along a cell diagonal, such as $(x, 2x, z)$, is up to $\sqrt{5}$ times longer than a cell edge, so a larger noise scale is needed for the wrapped noise distribution to become uniform along such a line.
At $\sqrt{5}/2$, the noise smears every atom uniformly along its free directions, so the fractional variant is corrupted to the same uniform state as the Cartesian model.
At 0.5, the noise is not uniform on sites whose degrees of freedom run along a cell diagonal.
In 4{,}000 MP-20 training structures, these sites are occupied by 9.1\% of the atoms and occur in 14.7\% of the structures.
DiffCSP++ keeps 0.5 and adds a Langevin corrector step at every reverse step.

\section{Atomic type diffusion details}
\label{si:types}
All atoms of a Wyckoff orbit must be of the same element for the crystal to be invariant under the operations of its space group.
We mask and unmask all atoms of an orbit together, so that the crystal retains its space group.
\citet{levy2025symmcd} and \citet{chang2025sgequidiff} use an equivalent formulation.
They sample the asymmetric unit and obtain the other atoms by applying the space-group operations.

Each orbit $o$ has one species variable $a^o_t \in \{1, \dots, K, m\}$ at step $t$, where $K = 100$ is the number of elements and $m$ is the mask state.
Training uses all $K$ elements.
During sampling, we exclude 24 elements by setting their logits to $-\infty$, which leaves 76.
The excluded elements are He, Ne, Ar, Kr, Xe, Tc, Pm and the 17 elements from Po to Fm.
MatterGen \citep{zeni2025mattergen} excludes the same elements during sampling.
About 6.7\% of MP-20 training structures contain an excluded element, mostly an actinide, Pm or Tc.
Alex-MP-20 contains none of them.
The symmetry-aware models of Table~\ref{tab:main} do not exclude these elements.
Between 6 and 8\% of their scored structures contain an excluded element.
%
At step $t$, an unmasked orbit is masked with probability $\beta_t$, and a masked orbit stays masked.
The single-step transition matrix of this absorbing-state kernel \citep{austin2021d3pm, zeni2025mattergen} is
\begin{equation}\label{si:eq:type_forward}
    [\mathbf{Q}_t]_{ij} = \begin{cases} 1 & i = j = m, \\ 1 - \beta_t & i = j \neq m, \\ \beta_t & j = m,\; i \neq m, \end{cases}
\end{equation}
and the probability that an orbit is still unmasked after $t$ steps falls linearly as $\bar\beta_t = (T - t)/T$, so that $\beta_t = 1 - \bar\beta_t/\bar\beta_{t-1} = 1/(T - t + 1)$.

The loss is the absorbing-state variational bound plus an auxiliary cross-entropy term with a small weight.
Only masked orbits contribute, since an unmasked orbit does not change state in the reverse process.
For each atom $i$ of a masked orbit $o(i)$, the bound term is the KL divergence between the posterior $q(a^{o(i)}_{t-1} \mid a^{o(i)}_t, a^{o(i)}_0)$ of the forward process and the model's reverse distribution.
The reverse distribution is parameterised by that atom's logits $\mathbf{u}_i$.
The type loss is
\begin{equation}\label{si:eq:type_loss}
    \mathcal{L}_{\mathrm{type}} = \mathbb{E}_t \Bigg[\, \frac{1}{N_m} \sum_{i : a^{o(i)}_t = m} \mathrm{KL}\big(q \,\|\, p_\theta\big)_i + \lambda_{\mathrm{ce}}\, \mathrm{CE}(\mathbf{u}_i, a^{o(i)}_0) \Bigg],
\end{equation}
where $N_m$ is the number of atoms in masked orbits.

\section{Training details}
\label{si:training}

\paragraph{Data.}
We convert each structure to its conventional standardised cell with spglib \citep{togo2024spglib} and assign its space group, Wyckoff letters and orbits with moyopy \citep{shinohara2026moyo}.
Both steps use a symmetry tolerance of 0.01~\AA{} and no angle tolerance.
The MP-20 split of \citet{xie2022cdvae} has 27,136 training, 9,047 validation and 9,046 test structures.
Alex-MP-20 \citep{zeni2025mattergen} is built from the Materials Project and Alexandria \citep{schmidt2022alexandria,schmidt2023alexandria}. It has 607,683 training and 67,521 validation structures and no test split.
Its structures have at most 20 atoms and an energy above the hull of at most 0.1~eV/atom.

\paragraph{Denoiser.}
The denoiser is an EquiformerV3 \citep{liao2026equiformerv3} with 13.7~million parameters.
It has four transformer blocks with 96 channels, eight attention heads and a feed-forward width of 512.
Its features use spherical harmonics up to degree $l = 3$.
The graph connects each atom to all atoms within a 7~\AA{} cutoff, including periodic images. Interatomic distances are expanded in 128 Gaussian radial basis functions.
Each atom has four inputs, all projected into its scalar ($l = 0$) features.
The first is its type, a distribution over the 100 elements and the mask state.
The second is the 231-dimensional Wyckoff embedding of Appendix~\ref{si:vae}, which is the only input that depends on the space group.
The third is the noise level of each of the three processes, encoded by a sinusoidal embedding of width 8.
The fourth is the self-conditioning input $\boldsymbol{\pi}$ \citep{chen2023analog}, the softmax of the type logits predicted for the atom at the previous step.
An extra entry is 1 when a prediction is present and 0 otherwise.
At the first sampling step, $\boldsymbol{\pi}$ and this entry are zero.
In training, a first pass of the denoiser without gradients predicts the types of the same noisy crystal.
For each crystal, $\boldsymbol{\pi}$ is set to this prediction with probability 0.5 and to zero otherwise.
The loss uses the output of a second pass.
Three linear heads give the outputs.
The type head maps each atom's scalar features to 100 logits, one per element.
During sampling, 24 of these logits are set to $-\infty$ (Appendix~\ref{si:types}).
The coordinate head maps each atom's vector ($l = 1$) features to one vector.
The lattice readout maps the $l = 0$ and $l = 2$ features to the six components of $\mathbf{S}$.

\paragraph{Optimisation.}
We train one diffusion model on MP-20 and one on Alex-MP-20 (Table~\ref{tab:training}).
Both minimise $20\,\mathcal{L}_{\mathrm{coord}} + 150\,\mathcal{L}_{\mathrm{type}} + 0.9\,\mathcal{L}_{\mathrm{lattice}}$.
The three terms differ in natural size (a squared error on the scaled coordinate target, a KL divergence over the 100 elements and a normalised error on the cell), so the weights mainly bring them to a comparable magnitude.
We then set the coordinate and lattice weights with a small grid search.
The auxiliary cross-entropy in the type loss of Equation~\ref{si:eq:type_loss} has weight $\lambda_{\mathrm{ce}} = 0.005$.
Training uses Adam with a learning rate of $10^{-3}$, a global batch of 128 crystals and gradient clipping at norm 1.
We compute the validation loss on the full validation set every three epochs. When it has not improved for 30 epochs, we reduce the learning rate by a factor of 0.6, down to a minimum of $10^{-4}$.
For sampling, we use an exponential moving average of the weights with decay 0.99, taken from the checkpoint with the lowest validation loss.
This checkpoint is used throughout the paper.

\begin{table}[H]
\centering
\caption{Diffusion training runs.}
\label{tab:training}
\small
\begin{tabular}{@{}lll@{}}
\toprule
 & MP-20 & Alex-MP-20 \\
\midrule
Epochs & 759 & 576 \\
GPUs & 4 H200 & 4 H200 \\
Global batch & 128 & 128 \\
Learning rate & $10^{-3}$ to $10^{-4}$ & $10^{-3}$ to $10^{-4}$ \\
Wall time per epoch & 2.6 min & 49.6 min \\
Wall time & 1.4 days & 19.8 days \\
\bottomrule
\end{tabular}
\end{table}

\section{Evaluation protocol}
\label{si:eval}

\subsection{LeMat-GenBench settings}
\label{si:lmb}
All benchmark metrics come from an unmodified version of LeMat-GenBench \citep{betala2025lematgenbench} at commit \texttt{58e6eae3}.
We run it with the three potentials ORB v3 (conservative, OMat24; \citealp{rhodes2025orbv3, barroso2026omat24}), MACE-MP (medium MPA-0; \citealp{batatia2025macemp}) and UMA-S-1 (OMat task; \citealp{wood2025uma}).
Each potential computes the energy of the structure as submitted.
The energy above the hull is the mean over the potentials that return a value.
A structure is valid if it is charge neutral, has a physically plausible density and cell, and has no two atoms closer than a limit set by their atomic radii.
Uniqueness and novelty are determined using the pymatgen structure matcher \citep{ong2013pymatgen}.
Novelty is checked against the 5,335,299 structures of LeMat-Bulk \citep{siron2025lematbulk}.

Every rate is a count divided by the 2,500 structures submitted.
To check our pipeline, we evaluated the released DiffCSP++ and SymmCD sets ourselves and reproduced every count in their leaderboard rows.

The benchmark does not count a stable structure as metastable.
A stable, unique and novel structure therefore counts towards the SUN rate and not the mSUN rate.
The SUN rate is 0.62\% averaged over the four unfiltered MP-20 sets and 2.12\% for the relaxed filtered set.
Every other set in this paper and every row of Tables~\ref{tab:main} and~\ref{tab:filter} has a SUN rate of at most 2.12\%.
Template filtering also lowered the stable fraction, from 4.97\% (mean over four sets) to 4.36\%.
The fall in the metastable fraction under filtering is therefore not a shift of structures into the stable class.

The benchmark relaxes each structure with each potential for 50 steps.
This relaxation is used only for the RMSD column, so we disabled it when computing the stability metrics.
We then re-evaluated the six MP-20 GEODE sets of Tables~\ref{tab:main} and~\ref{tab:filter} with it enabled to obtain their RMSD values.
These re-evaluations reproduced every mSUN count exactly and the mean energies above the hull to within $10^{-8}$~eV/atom.
The Alex-MP-20 GEODE sets were evaluated once, with the relaxation enabled.

\subsection{Pre-relaxation}
\label{si:prerelax}
Results marked with an asterisk come from structures relaxed before submission with the EquiformerV3 potential of Section~\ref{sec:setup}.
We relax the atomic positions and the cell with FIRE \citep{bitzek2006fire} and a Fr\'echet cell filter until the maximum force is below 0.02~eV/\AA{}, for at most 500 steps.
For a structure that has not converged after 500 steps, we use its last geometry.
We then evaluate the relaxed structures in the same way as the unrelaxed ones.

After pre-relaxation, the benchmark RMSD of the MP-20 sets is 0.151~\AA{} unfiltered and 0.203~\AA{} filtered, compared with 0.42 to 0.51~\AA{} as generated.
The benchmark relaxes with its own three potentials, not with EquiformerV3.
The remaining RMSD can therefore reflect differences between the minimum-energy geometries of these potentials and of EquiformerV3.

\subsection{Symmetry detection}
Figure~\ref{fig:spg} and the fidelity check of Section~\ref{sec:spg_results} use the space group and point group of each generated structure.
We detect both with moyopy on the conventional cell, at a symmetry tolerance of 0.01~\AA{} and no angle tolerance, as for the training data (Appendix~\ref{si:training}).
We generated the MatterGen structures of Figure~\ref{fig:spg} from the released MP-20 checkpoint and detected their symmetry in the same way.

\subsection{Symmetry on Alex-MP-20}
\label{si:alex_pg}
\begin{figure}[t]
    \centering
    \includegraphics[width=\textwidth]{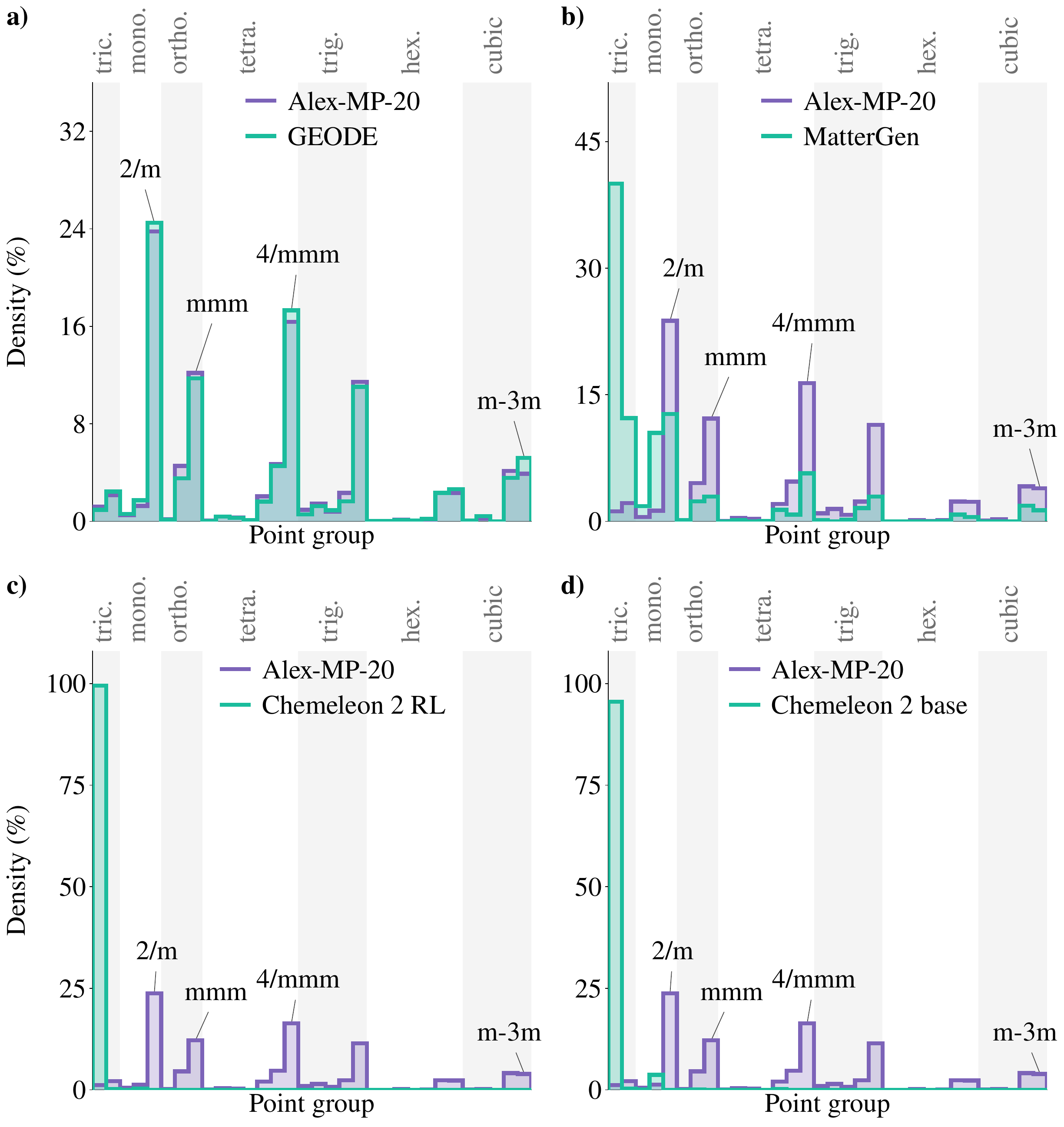}
    \caption{Point-group distribution of \textbf{(a)} GEODE, \textbf{(b)} MatterGen, \textbf{(c)} Chemeleon~2 after reinforcement-learning fine-tuning and \textbf{(d)} the Chemeleon~2 base model, compared with the Alex-MP-20 training set. Point groups detected on unrelaxed structures at a symmetry tolerance of 0.01~\AA{}. The 32 point groups are ordered by increasing symmetry, with shaded bands indicating the crystal systems.}
    \label{fig:alex_pg}
\end{figure}

On Alex-MP-20, the point-group distribution of GEODE is closer to that of the training set than those of MatterGen and Chemeleon~2 (Figure~\ref{fig:alex_pg}).
We generated 2{,}500 MatterGen structures from the released Alex-MP-20 checkpoint.
We used the released Chemeleon~2 base checkpoint (before reinforcement-learning fine-tuning) \citep{park2026chemeleon2} and generated 2{,}500 structures from it.
For Chemeleon~2 after fine-tuning, we use the first 2{,}500 of its 10{,}000 released structures.
Point groups are detected as above.
Of the 10{,}000 GEODE structures, 0.94\% are assigned to $P1$, compared with 1.20\% of the training structures, 40.00\% for MatterGen, 95.52\% for the Chemeleon~2 base model and 99.48\% for Chemeleon~2 after fine-tuning.
The Jensen--Shannon distance between the point-group distribution of GEODE and that of the training set is 0.06, compared with 0.59 for MatterGen, 0.96 for the Chemeleon~2 base model and 0.97 for Chemeleon~2 after fine-tuning.

\section{Choosing the template filter}
\label{si:template_rationale}

The template filter of Section~\ref{sec:filter_results} accepts a sampled template only if two counts fall within specified ranges.
The first is the number of occupied Wyckoff orbits in the template.
The second is its training-set frequency, the number of training crystals with the same template.
Two crystals share a template when they have the same space group and the same number of distinct orbits on each Wyckoff letter.
We obtain the templates of the training crystals from the moyopy labels described in Appendix~\ref{si:training}.
The MP-20 and Alex-MP-20 training sets contain 3,358 and 5,838 distinct templates.
We count frequency in the training set rather than in LeMat-Bulk, the reference set for the novelty metric.
Filtering directly on the novelty reference set would make the evaluation dependent on the selection criterion.

To choose the ranges, we generated 10{,}000 crystals from unfiltered templates for each training set, as four sets of 2{,}500.
We evaluated them using the main-text protocol and computed mSUN for each orbit count and training-set frequency range (Table~\ref{tab:mp20_orbit_bands}).

\begin{table}[H]
\centering
\caption{mSUN of the 10{,}000 structures generated from unfiltered templates for each training set, grouped by the number of occupied Wyckoff orbits and by training-set frequency.
Novel and mSUN are percentages of all structures submitted, while Metastable is the percentage of novel structures that are metastable.
The mSUN percentage is therefore the product of the Novel and Metastable percentages divided by 100.
No MP-20 template occurs more than 2{,}525 times.}
\label{tab:mp20_orbit_bands}\label{tab:mp20_rarity_bands}\label{tab:alex_orbit_bands}\label{tab:alex_rarity_bands}
\small
\resizebox{\textwidth}{!}{%
\begin{tabular}{@{}l rrrr rrrr@{}}
\toprule
& \multicolumn{4}{c}{MP-20} & \multicolumn{4}{c}{Alex-MP-20} \\
\cmidrule(lr){2-5} \cmidrule(l){6-9}
& & Novel & Metastable & mSUN & & Novel & Metastable & mSUN \\
& Structures & (\%) & among novel (\%) & (\%) & Structures & (\%) & among novel (\%) & (\%) \\
\midrule
\multicolumn{9}{@{}l}{\emph{Occupied orbits}} \\
1 & 121 & 9.9 & 41.7 & 4.13 & 24 & 29.2 & 14.3 & 4.17 \\
2 & 1204 & 28.4 & 31.0 & 8.80 & 469 & 34.5 & 45.7 & 15.78 \\
3 & 3235 & 32.3 & 28.2 & 9.12 & 2357 & 35.6 & 45.1 & 16.04 \\
4 & 2298 & 60.6 & 25.2 & 15.27 & 2716 & 56.5 & 49.1 & 27.76 \\
5 & 1132 & 81.8 & 21.1 & 17.23 & 1617 & 65.2 & 51.2 & 33.40 \\
6 & 769 & 87.6 & 19.4 & 17.04 & 1716 & 59.8 & 53.6 & 32.05 \\
7 or more & 1241 & 95.9 & 9.2 & 8.78 & 1101 & 71.4 & 35.8 & 25.52 \\
\midrule
\multicolumn{9}{@{}l}{\emph{Occurrences in the training set}} \\
0 & 892 & 89.7 & 9.9 & 8.86 & 336 & 88.1 & 13.2 & 11.61 \\
1 to 10 & 3289 & 77.2 & 18.9 & 14.56 & 527 & 79.9 & 28.0 & 22.39 \\
11 to 100 & 3293 & 55.7 & 26.2 & 14.58 & 1235 & 73.7 & 36.9 & 27.21 \\
101 to 1{,}000 & 1289 & 28.2 & 37.6 & 10.63 & 2777 & 59.3 & 47.2 & 28.02 \\
1{,}001 to 5{,}000 & 1237 & 3.7 & 37.0 & 1.37 & 2372 & 52.9 & 57.5 & 30.40 \\
above 5{,}000 & -- & -- & -- & -- & 2753 & 32.0 & 66.6 & 21.29 \\
\midrule
all templates & 10000 & 55.8 & 21.4 & 11.92 & 10000 & 54.1 & 47.7 & 25.78 \\
\bottomrule
\end{tabular}}
\end{table}

Novelty generally rises as the number of orbits increases and as training-set frequency decreases.
On MP-20, the metastable fraction among novel structures decreases with the number of orbits and generally increases with frequency.
The mSUN rate therefore peaks at intermediate values: 5 orbits and 11--100 occurrences, with a similar rate at 1--10 occurrences.
On Alex-MP-20, the metastable fraction among novel structures instead rises with the number of orbits and drops only for templates with seven or more orbits, yet mSUN still peaks at 5 orbits.
By frequency, Alex-MP-20 shows the same trends as MP-20, but mSUN peaks at 1{,}001--5{,}000 occurrences.
Alex-MP-20 is over twenty times larger than MP-20, so a template that occurs 1{,}001 times makes up a much smaller fraction of its training set.

We selected ranges of 4--6 orbits for both datasets, with 1--1{,}000 training-set occurrences for MP-20 and 11--5{,}000 for Alex-MP-20.
These ranges balance estimated mSUN against the fraction of templates retained.
Some narrower ranges give a slightly higher estimated mSUN but retain fewer templates (Table~\ref{tab:mp20_filters}).
The RMSD of the as-generated structures increases as the template filter becomes stricter (Table~\ref{tab:filter}).
Stricter filters therefore retain structures that undergo larger atomic displacements during relaxation, yet the filtered sets have a higher mSUN rate.

To generate a filtered set, we sample templates from the template VAE and accept each sample whose two counts fall within the chosen ranges.
We keep duplicates, so the accepted templates follow the prior conditioned on satisfying the filter.
For the filtered sets of the main text, the filter accepted 2,754 of 8,192 samples on MP-20 (33.6\%) and 2,705 of 7,168 on Alex-MP-20 (37.7\%).
We generated structures from the accepted templates in the order sampled and evaluated the first 2,500.
On these new sets, the measured mSUN is 0.3 percentage points below the estimate from the unfiltered structures on MP-20 and 1.9 percentage points below it on Alex-MP-20, compared with a sampling error of about 0.9 percentage points for a set of 2{,}500.

\begin{table}[H]
\centering
\small
\setlength{\tabcolsep}{3.2pt}
\renewcommand{\arraystretch}{1.08}
\caption{Effect of template filtering on MP-20 generation.
Acceptance is the percentage of sampled templates retained, counted over templates.
The columns ``SUN, relaxed'' and ``mSUN, relaxed'' report SUN and mSUN after relaxation with EquiformerV3.
In the no filters row, Novel, Metastable, SUN and mSUN are means over four sets of 2{,}500 structures. All other values are from one set.}
\vspace{0.5\baselineskip}
\label{tab:filter}
\begin{tabular}{l cccccccc}
\toprule
Filter & Acceptance & Novel & Metastable & SUN & mSUN & SUN, relaxed & mSUN, relaxed & RMSD \\
& (\%) & (\%) $\uparrow$ & (\%) $\uparrow$ & (\%) $\uparrow$ & (\%) $\uparrow$ & (\%) $\uparrow$ & (\%) $\uparrow$ & (\AA) $\downarrow$ \\
\midrule
None & 100.0 & 55.8 & 41.9 & 0.62 & 11.92 & 1.20 & 16.40 & 0.419 \\
Frequency & 78.0 & 60.2 & 40.6 & 0.76 & 13.76 & 1.44 & 18.84 & 0.458 \\
Orbit count & 41.9 & 72.1 & 35.8 & 1.08 & 16.40 & 1.96 & 21.56 & 0.512 \\
Both & 33.6 & 74.6 & 35.1 & 1.20 & 18.20 & 2.12 & 23.80 & 0.514 \\
\bottomrule
\end{tabular}%
\end{table}

\begin{table}[H]
\centering
\caption{Candidate filters.
Share of output is the percentage of the 10{,}000 unfiltered structures whose templates pass the filter, counted over structures.
Estimated mSUN is the rate among those structures.
Measured mSUN is the rate among 2{,}500 new structures generated with the filter, where such a set exists.
For the rows without a filter, it is the mean over the four unfiltered sets, which are the structures used for the estimate.
The bold rows are the filters used in the main text.
The last row applies the MP-20 frequency range to Alex-MP-20 unchanged.}
\label{tab:mp20_filters}\label{tab:alex_filters}
\small
\begin{tabular}{@{}llrrr@{}}
\toprule
& & Share of & Estimated & Measured \\
Orbits & Occurrences & output (\%) & mSUN (\%) & mSUN (\%) \\
\midrule
\multicolumn{5}{@{}l}{\emph{MP-20}} \\
all & all & 100.0 & 11.92 & 11.92 \\
4 to 6 & all & 42.0 & 16.12 & 16.40 \\
all & 1 to 1{,}000 & 78.7 & 13.92 & 13.76 \\
\textbf{4 to 6} & \textbf{1 to 1{,}000} & \textbf{33.5} & \textbf{18.50} & \textbf{18.20} \\
4 to 5 & 1 to 1{,}000 & 27.3 & 18.39 & -- \\
5 to 6 & 1 to 1{,}000 & 16.1 & 18.62 & -- \\
4 to 6 & 11 to 1{,}000 & 16.6 & 20.00 & -- \\
4 to 6 & 1 to 100 & 30.8 & 17.98 & -- \\
\midrule
\multicolumn{5}{@{}l}{\emph{Alex-MP-20}} \\
all & all & 100.0 & 25.78 & 25.78 \\
4 to 6 & all & 60.5 & 30.48 & -- \\
all & 11 to 5{,}000 & 63.8 & 28.74 & -- \\
\textbf{4 to 6} & \textbf{11 to 5{,}000} & \textbf{39.4} & \textbf{33.78} & \textbf{31.92} \\
4 to 6 & 101 to 5{,}000 & 33.1 & 34.12 & -- \\
5 to 6 & 11 to 5{,}000 & 20.5 & 35.22 & -- \\
4 to 6 & 1{,}001 to 5{,}000 & 17.3 & 34.14 & -- \\
4 to 6 & 1 to 1{,}000 & 24.7 & 32.47 & -- \\
\bottomrule
\end{tabular}
\end{table}

\section{Property steering details}
\label{si:dielectric}

For property steering (Section~\ref{sec:dielectric_results}), we fine-tune the MP-20 diffusion model on permittivity labels and generate cubic crystals at several targets and guidance strengths.
We use the AnisoNet training set \citep{lou_discovery_2025}, which includes all crystal systems.
Each crystal has a high-frequency (electronic) dielectric tensor computed with density functional perturbation theory.
We define its polycrystalline permittivity as the mean eigenvalue of the symmetrised tensor and use the logarithm of this value as the training label.
The labels therefore come from first-principles calculations.
AnisoNet is used only to predict the permittivity of generated crystals.
Before taking the logarithm, the permittivities have a median of 4.01, a 90th percentile of 8.39 and a maximum of 14.89.
The dataset excludes values above 15, placing the highest generation target just beyond the observed label range.

We standardise the logarithmic labels using the full dataset and encode them with 96 fixed sinusoidal features.
FiLM layers \citep{perez2018film} after each of the four denoiser blocks scale and shift the scalar features using affine functions of this encoding.
Starting from the final unconditional MP-20 checkpoint, we train only the FiLM parameters, accounting for 74,496 of the model's 13.7~million parameters.
Their zero initialisation leaves the features unchanged, so the model initially reproduces the unconditional model.
We fine-tune using Adam with a learning rate of $10^{-4}$ and a global batch size of 128 for 195 epochs on RTX 6000 GPUs, taking 6.2~h in total.
We use the final parameters without weight averaging.

\begin{figure}[H]
    \centering
    \includegraphics[width=\textwidth]{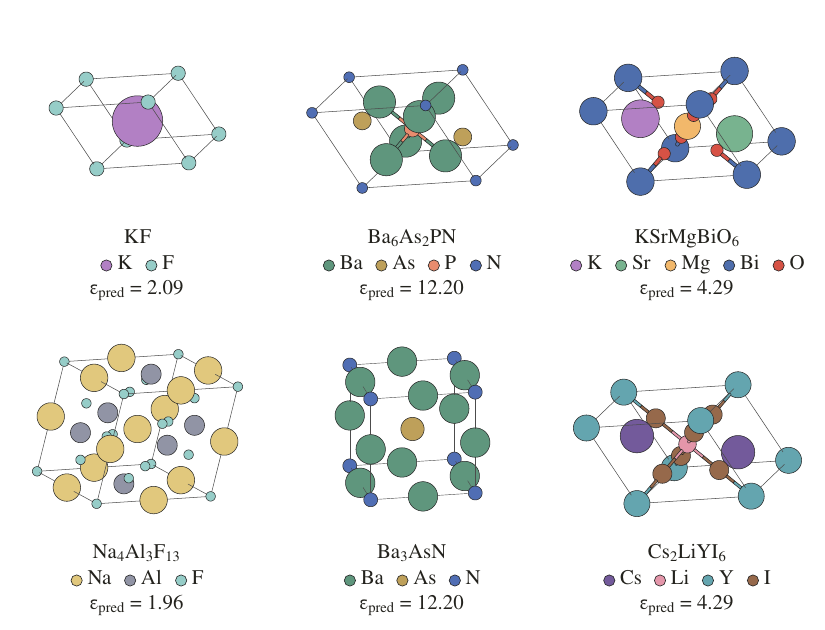}
    \caption{Example cubic crystals generated at a low permittivity target (left), at a high target (middle) and without a target (right).
    Each structure is labelled with its formula and AnisoNet-predicted permittivity.
    Examples were selected to lie within 0.1 of the median predicted permittivity of their respective generated sets.
    The low-target examples are fluorides of light cations.
    The high-target examples are barium pnictides, including the antiperovskite Ba$_3$AsN.
    The control examples are halide and oxide double perovskites.}
    \label{fig:diel_examples}
\end{figure}

The template VAE produces cubic templates in about a quarter of samples.
We generated structures for every combination of target $\varepsilon \in \{2, 5, 10, 15\}$ and guidance strength $\gamma_{\mathrm{g}} \in \{1, 2, 3, 5\}$.
Every combination uses the same 7,500 cubic templates in the same order, one structure per template, until about 2,500 pass the screen.
We also used the same fine-tuned model to generate a control set without a target.
AnisoNet predicts the permittivity of each unrelaxed structure.
By default, its predictions depend on a normalisation constant computed from the input set of structures.
We fix this constant at its training-set value so that predictions are comparable across all generated sets.
At $\gamma_{\mathrm{g}} = 5$, fluorides account for 95.3\% of the structures classified as insulating at a target of 2.
At higher targets, the largest composition classes are oxides at 5, nitrides at 10, and pnictides and other compounds at 15 (Table~\ref{tab:diel_chemistry}).
Figure~\ref{fig:diel_examples} shows example structures.

\begin{table}[H]
\centering
\caption{Composition statistics for structures retained by the insulating-composition screen, shown as percentages of each generated set.
Target-conditioned sets use $\gamma_{\mathrm{g}} = 5$; the control set is generated without a property target.
Composition classes are assigned according to the most electronegative element, and a structure whose most electronegative element has a Pauling electronegativity below 2.0 has no anion.
Charge balancing is possible if some choice of one oxidation state per element from the ICSD oxidation-state table gives zero total charge.
The final three rows report additional statistics and overlap with the composition classes.}
\label{tab:diel_chemistry}
\small
\begin{tabular}{@{}lrrrrr@{}}
\toprule
 & Control & $\varepsilon = 2$ & $\varepsilon = 5$ & $\varepsilon = 10$ & $\varepsilon = 15$ \\
\midrule
Fluorides & 22.1 & 95.3 & 0.1 & 0.0 & 0.1 \\
Oxides & 24.8 & 1.2 & 48.2 & 2.5 & 0.5 \\
Other halides & 22.7 & 1.4 & 17.5 & 6.7 & 2.3 \\
Chalcogenides (S, Se, Te) & 8.9 & 0.0 & 7.8 & 17.1 & 11.9 \\
Nitrides & 5.0 & 0.0 & 23.2 & 43.8 & 22.9 \\
Other pnictides and other compounds & 15.4 & 1.8 & 3.2 & 29.4 & 60.6 \\
No anion & 1.1 & 0.2 & 0.0 & 0.6 & 1.7 \\
\midrule
Contains Ti, Zr, Hf, Nb or Ta & 7.7 & 0.0 & 21.2 & 13.1 & 2.6 \\
Contains Ba & 9.9 & 2.9 & 26.8 & 35.7 & 33.7 \\
Charge balance possible & 80.4 & 34.4 & 72.9 & 73.8 & 79.9 \\
\bottomrule
\end{tabular}
\end{table}

\subsection{Composition screening}
\label{si:metal_screen}

The composition screening of Section~\ref{sec:dielectric_results} classifies each generated crystal as likely metallic or insulating.
It combines the ICSD oxidation-state table of LeMat-GenBench \citep{betala2025lematgenbench} with the SMACT metallicity score \citep{davies2019smact}, using a threshold of 0.60.
We apply this screening after generation.
It estimates metallicity but does not enforce charge neutrality, so a structure can pass the screen and still fail the benchmark's charge-neutrality check.

We generated structures in batches until approximately 2,500 were classified as insulating for each combination of target and guidance strength, and evaluated all retained structures.
At $\gamma_{\mathrm{g}} = 5$, the retained sets contain between 2,493 and 2,654 structures, with 2,496 in the control set.
Table~\ref{tab:dielectric_si} reports the median predicted permittivity, the fraction classified as metallic and the validity of the retained structures for each combination.
At $\gamma_{\mathrm{g}} = 5$, the fraction classified as metallic rises from 0.1\% at a target of 2 to 46.3\% at a target of 15, compared with 54.9\% for the control.

\begin{table}[H]
\centering
\caption{Property-guided generation at different targets and guidance strengths.
Rows specify the guidance strength $\gamma_{\mathrm{g}}$, and columns within each group specify the target permittivity $\varepsilon$.
The three groups report the median AnisoNet-predicted permittivity of structures classified as insulating, the percentage of all generated structures classified as metallic, and the percentage of retained structures that pass the validity checks of Section~\ref{sec:setup}.
The control set is generated without a property target.}
\label{tab:dielectric_si}\label{tab:diel_validity}
\small
\setlength{\tabcolsep}{3.5pt}
\begin{tabular}{@{}l cccc cccc cccc@{}}
\toprule
& \multicolumn{4}{c}{Median predicted $\varepsilon$} & \multicolumn{4}{c}{Classified metallic (\%)} & \multicolumn{4}{c}{Validity (\%)} \\
\cmidrule(lr){2-5} \cmidrule(lr){6-9} \cmidrule(l){10-13}
$\gamma_{\mathrm{g}}$ & 2 & 5 & 10 & 15 & 2 & 5 & 10 & 15 & 2 & 5 & 10 & 15 \\
\midrule
1 & 2.87 & 4.91 & 5.61 & 5.67 & 30.7 & 52.3 & 54.3 & 55.3 & 74.7 & 89.4 & 89.6 & 90.1 \\
2 & 2.43 & 5.26 & 7.22 & 8.69 & 3.6 & 43.8 & 50.8 & 52.2 & 57.8 & 88.0 & 87.8 & 86.7 \\
3 & 2.19 & 5.51 & 9.31 & 11.20 & 0.4 & 30.8 & 45.8 & 48.8 & 53.2 & 85.3 & 83.9 & 86.6 \\
5 & 2.05 & 6.22 & 10.47 & 12.19 & 0.1 & 18.6 & 41.3 & 46.3 & 60.1 & 88.4 & 82.7 & 88.1 \\
\midrule
Control & \multicolumn{4}{c}{4.29} & \multicolumn{4}{c}{54.9} & \multicolumn{4}{c}{90.4} \\
\bottomrule
\end{tabular}
\end{table}

\subsection{Guidance strength and validity}
\label{si:diel_grid}
As $\gamma_{\mathrm{g}}$ decreases from 5 to 3, 2 and 1, the median predicted permittivity for each target moves towards the control median of 4.29 (Table~\ref{tab:dielectric_si}).
At a target of 2, 74.7\% of structures classified as insulating pass the validity checks at $\gamma_{\mathrm{g}} = 1$, compared with 53.2--60.1\% at stronger guidance.
At the other targets, validity ranges from 82.7\% to 90.1\%, compared with 90.4\% for the control.
At a target of 2 and $\gamma_{\mathrm{g}} \geq 3$, charge balance is possible for only about one third of the screened structures, compared with 72--80\% at targets of 5 and above.
This suggests that failure to satisfy charge neutrality contributes to the lower validity at a target of 2.
Every charge-balanced structure also passes the benchmark's looser charge-neutrality check.

\section{Symmetry of the denoiser output}
\label{si:equivariance}

The denoiser is a graph neural network that is invariant under translations and equivariant under rotations of its input.
When the input crystal is rotated, the predicted type logits are unchanged, the predicted coordinate targets rotate with the atoms, and the predicted deformation tensor transforms as a rank-2 tensor.
We call a space-group operation proper if its orthogonal part is a rotation, that is, if it involves no reflection, inversion or rotoinversion.
Formally, let $g$ be a proper space-group operation with orthogonal part $\mathbf{O}_g$, and let $gi$ be the atom to which $g$ maps atom $i$.
For any crystal $C$, the predictions satisfy
\begin{equation}\label{si:eq:equivariance}
    \mathbf{u}_{gi}(gC) = \mathbf{u}_i(C), \qquad
    \hat{\mathbf{s}}_{gi}(gC) = \mathbf{O}_g\, \hat{\mathbf{s}}_i(C), \qquad
    \mathbf{S}(gC) = \mathbf{O}_g\, \mathbf{S}(C)\, \mathbf{O}_g^{\top},
\end{equation}
where $\mathbf{u}_i$ is the vector of type logits of atom $i$, $\hat{\mathbf{s}}_i$ is its scaled coordinate target and $\mathbf{S}$ is the deformation tensor.
A noisy crystal that is invariant under $g$ satisfies $gC = C$.
Substituting $gC = C$ into Equation~\ref{si:eq:equivariance} gives
\begin{equation}\label{si:eq:fixed_point}
    \mathbf{u}_{gi} = \mathbf{u}_i, \qquad
    \hat{\mathbf{s}}_{gi} = \mathbf{O}_g\, \hat{\mathbf{s}}_i, \qquad
    \mathbf{S} = \mathbf{O}_g\, \mathbf{S}\, \mathbf{O}_g^{\top}.
\end{equation}
Atoms related by $g$ therefore have equal type logits, and their coordinate targets are related by $\mathbf{O}_g$.
If $g$ is not the identity and maps atom $i$ onto itself, the second condition restricts the target to the rotation axis of $g$.
The third condition is Neumann's principle for the deformation tensor.
These constraints follow from the architecture and hold for any network weights.

The argument needs two conditions.
First, the input must be invariant under the space group.
We therefore project the coordinate noise onto the free directions of each Wyckoff position (Section~\ref{sec:coord_diffusion}) and the lattice noise onto the subspace allowed by the crystal system (Section~\ref{sec:lattice_diffusion}).
Second, the network must be equivariant under every operation of the space group.
The EquiformerV3 denoiser is equivariant under rotations but not under reflections or inversion \citep{liao2026equiformerv3}.
For the 65 space groups that contain only proper operations, Equation~\ref{si:eq:fixed_point} holds for all operations.
For the other 165 space groups, it holds only for the proper operations.
During sampling, we symmetrise the predicted coordinate targets with $\mathcal{R}_G$, average the type logits within each Wyckoff orbit and constrain the lattice with $\mathcal{P}_G$ for every space group.
For these 165 groups, the three steps impose the constraints of the improper operations.
For a symmetric rank-2 tensor, the rotations of the point group already impose the constraints of the full point group, except for the point groups $m$, $mm2$, $\bar{4}$ and $\bar{4}2m$.
With a network that is also equivariant under reflections and inversion, these projections would only remove floating-point error.

During generation, the network computes features only for one atom of each set of atoms related by proper operations and obtains the features of the other atoms with the Wigner-D matrices of $\mathbf{O}_g$.
This set of atoms is the asymmetric unit for the 65 groups above and contains at most twice as many atoms for the other groups.

\section{Generation cost}
\label{si:cost}

GEODE generates 2{,}500 MP-20 structures in 2.16~h on one Quadro RTX 6000 GPU with 24~GB of memory (Table~\ref{tab:cost}).
This timing uses evaluation on the asymmetric unit (Appendix~\ref{si:equivariance}).
Template sampling takes 15~ms per template on a GPU for the template VAE.
The template filter of Section~\ref{sec:filter_results} rejects about two thirds of sampled templates.
On unconditional generation of the MP-20 model, obtaining templates for 2{,}500 structures therefore takes about 30 seconds, compared with 2.16~h for diffusion sampling.

\begin{table}[H]
\centering
\caption{Generation time for 2{,}500 structures on one Quadro RTX 6000 GPU, using each model's released code and default sampler, and for GEODE evaluation on the asymmetric unit.
Timings exclude template sampling.
SGFM used a batch size of 1.}
\label{tab:cost}
\begin{tabular}{lrr}
\toprule
Model & 2{,}500 structures & Per structure \\
\midrule
Crystalite~\citep{veljkovic2026crystalite} & 3.5 min & 0.08 s \\
SGEquiDiff~\citep{chang2025sgequidiff} & 26.6 min & 0.64 s \\
SGFM~\citep{puny2025sgfm} & 80.6 min & 1.93 s \\
MatterGen~\citep{zeni2025mattergen} & 6.06 h & 8.7 s \\
\midrule
\textbf{GEODE} & \textbf{2.16 h} & \textbf{3.1 s} \\
\bottomrule
\end{tabular}
\end{table}

\end{document}